\documentclass[a4paper,fleqn,usenatbib,useAMS]{mnras}

\usepackage{graphicx}	% Including figure files
\usepackage{amsmath}	% Advanced maths commands
\usepackage{amssymb}	% Extra maths symbols
\usepackage{multicol}   % Multi-column entries in tables
\usepackage{bm}		    % Bold maths symbols, including upright Greek
\usepackage{pdflscape}	% Landscape pages

\usepackage[T1]{fontenc}
\usepackage{ae,aecompl}
\usepackage{newtxtext,newtxmath}

\newcommand{\AstronomyLetters}{Astron. Lett.}
\newcommand{\AstronomyReports}{Astron. Rep.}  %!

\newcommand{\EuropeanPhysicalJournalD}{Eur. Phys. J. D}

\newcommand{\InternationalReviewsInPhysicalChemistry}{Int. Rev. Phys. Chem.}
\newcommand{\JournalOfMolecularSpectroscopy}{J. Mol. Spectrosc.}
\newcommand{\JournalOfTheAmericanChemicalSociety}{J. Am. Chem. Soc.}
\newcommand{\JournalOfPhysicalChemicalReferenceData}{J. Phys. Chem. Ref. Data}
\newcommand{\JournalOfChemicalPhysics}{J. Chem. Phys.}
\newcommand{\JournalOfPhysicsConferenceSeries}{J. Phys. Conf. Ser.}

\newcommand{\PlasmaPhysicsAndControlledFusion}{Plasma Phys. Contr. F.}
\newcommand{\ProceedingsOfTheNationalAcademyOfScience}{P. Natl. Acad. Sci.}

\newcommand{\ResearchInAstronomyAndAstrophysics}{Res. Astron. Astrophys.}

\title[Modelling cosmic masers in C-type shock waves]{Modelling cosmic masers in C-type shock waves -- the coexistence of Class I CH$_3$OH and 1720~MHz OH masers}

\author[A. V. Nesterenok] {A.V. Nesterenok,$^{1}$\thanks{E-mail:alex-n10@yandex.ru} \\
$^{1}$ Ioffe Institute, 26 Polytechnicheskaya St., 194021 Saint Petersburg, Russia \\
}
\date{Accepted XXX. Received YYY; in original form ZZZ}
\pubyear{2019}

\begin{document}
\label{firstpage}
\pagerange{\pageref{firstpage}--\pageref{lastpage}}
\maketitle

\begin{abstract}
The collisional pumping of CH$_3$OH and OH masers in non-dissociative C-type shock waves is studied. The chemical processes responsible for the evolution of molecule abundances in the shock wave are considered in detail. The large velocity gradient approximation is used to model radiative transfer in molecular lines. We present calculations of the optical depth in maser transitions of CH$_3$OH and OH for a grid of C-type shock models that vary in cosmic ray ionization rate, gas density and shock speed. We show that pre-shock gas densities $n_\text{H,tot} = 2 \times 10^4$--$2 \times 10^5$~cm$^{-3}$ are optimal for pumping of methanol maser transitions. A complete collisional dissociation of methanol at the shock front takes place for shock speeds $u_\text{s} \gtrsim 25$~km~s$^{-1}$. At high pre-shock gas density $n_\text{H,tot} = 2 \times 10^{6}$~cm$^{-3}$, the collisional dissociation of methanol takes place at shock speeds just above the threshold speed $u_\text{s} \approx$ 15--17.5~km~s$^{-1}$ corresponding to sputtering of icy mantles of dust grains. We show that the methanol maser transition E~$4_{-1} \to 3_0$ at 36.2~GHz has the optical depth $\vert \tau \vert$ higher than that of the transition A$^+$~$7_0 \to 6_1$ at 44.1~GHz at high cosmic ray ionization rate $\zeta_\mathrm{H_2} \gtrsim 10^{-15}$~s$^{-1}$ and pre-shock gas density $n_\text{H,tot} = 2 \times 10^4$~cm$^{-3}$. These results can be applied to the interpretation of observational data on methanol masers near supernova remnants and in molecular clouds of the Central Molecular Zone. At the same time, a necessary condition for the operation of 1720~MHz OH masers is a high ionization rate of molecular gas, $\zeta_\mathrm{H_2} \gtrsim 10^{-15}$~s$^{-1}$. We find that physical conditions conducive to the operation of both hydroxyl and methanol masers are cosmic ray ionization rate $\zeta_\mathrm{H_2} \approx 10^{-15}$--$3 \times 10^{-15}$~s$^{-1}$, and a narrow range of shock speeds $15 \lesssim u_\text{s} \lesssim 20$~km~s$^{-1}$. The simultaneous observations of OH and CH$_3$OH masers may provide restrictions on the physical parameters of the interstellar medium in the vicinity of supernova remnants. 
\end{abstract}


\section{Introduction}
Shock waves in the interstellar medium are typically caused by supernova explosions, outflows from young stellar objects, collisions between molecular clouds. The shock wave propagation in a molecular cloud can be traced by intense emission in atomic and molecular lines including the maser emission in a variety of species. Here, we focus on magnetohydrodynamic non-dissociative shock waves -- C-type shocks -- and on the generation of maser emission in such shocks. Observations of spectral lines of various molecules provide a tool to derive physical parameters in the interstellar gas, and we consider the maser emission of CH$_3$OH and OH molecules. 

Methanol masers have been empirically divided into two varieties, termed Class I and II \citep{Batrla1987,Menten1991b}. The Class II methanol masers reside in close proximity to individual high-mass young stellar objects, and these masers have a radiative pumping mechanism \citep[e.g.,][]{Cragg1992}. In contrast, the pumping mechanism of Class I masers is collisional. The Class I methanol masers are found in regions of both high- and low-mass star formation, and often offset ($\sim$ 0.1--1~pc) from radio-continuum and infrared sources \citep{Kurtz2004,Voronkov2014}. These masers trace shocked gas in star-forming regions, and can be associated with multiple phases in the evolution of a young stellar object such as outflows and expanding \ion{H}{II} regions \citep{Voronkov2014,Breen2019}. Class I methanol masers are also observed in the vicinity of supernova remnants (SNRs) \citep{Pihlstrom2014}, and in the Galactic Centre region, where they may be born in cloud--cloud collision shocks \citep{Sobolev1992,Salii2002,Zeng2020}. Within the Milky Way, the most widespread Class I methanol transitions are 44.1~GHz A$^+$~$7_0 \to 6_1$ and 95.2~GHz A$^+$~$8_0 \to 7_1$, other common transitions are 36.2~GHz E~$4_{-1} \to 3_0$ and 84.5~GHz E~$5_{-1} \to 4_0$ \citep[e.g.,][]{Ladeyschikov2019}. 

Maser emission of the hydroxyl radical is observed for all transitions of the ground rotational state of the molecule $^2\Pi_{3/2}$ $j = 3/2$ -- main lines 1665 and 1667 MHz and satellite lines 1612 and 1720~MHz. Generally, the maser emission in the 1665 and 1667~MHz transitions is associated with compact \ion{H}{II} regions, while masers in the 1612~MHz transition are produced in the envelopes of evolved stars \citep[e.g.,][]{Beuther2019,Qiao2020}. The 1720~MHz masers are sometimes found in star-forming regions and seldom in evolved stars. All these masers need a radiation field as a pumping source. But there is a class of relatively rare masers in the 1720~MHz OH transition associated with SNRs \citep{Wardle2002}. These masers are collisionally pumped and are produced in the warm molecular gas behind the supernova shock. The signature of collisional pumping is that other lines of the ground OH rotational state are generally detected in absorption \citep{Hewitt2008}. The collisional excitation of OH molecule in SNRs and the emergence of maser radiation were studied in our recent paper \citep{Nesterenok2020}. 

The collapse of massive stars often takes place in the vicinity of molecular clouds, and the supernova shock encounters and propagates through it. The hydroxyl masers 1720~MHz are considered as a tracer of interaction between a SNR and a neighbouring molecular cloud \citep{Pastchenko1974,Frail1996,Frail1998}. The OH maser emission at 1720~MHz has been detected in about 10 per cent of SNRs in our Galaxy \citep{Brogan2013}. One can expect that Class I methanol masers could be another signpost of the SNR--molecular-cloud interaction as these masers also trace shocks and have a collisional pumping mechanism. The methanol masers were found in the vicinity of the Sgr~A East SNR in the Galactic Centre \citep{Szczepanski1989,Sjouwerman2010,Pihlstrom2011}. Targeted searches for methanol masers towards SNRs in which OH maser emission had been observed were carried out \citep{Litovchenko2011,Pihlstrom2014}. It was found that the presence of OH emission in a SNR is not a sufficient condition for the detection of Class I methanol masers. \citet{Pihlstrom2014} searched for 36.2 and 44.1~GHz methanol lines in a sample of 21 Galactic SNRs, and reported the detection of methanol emission in two SNRs G1.4-0.1 and W28. \citet{McEwen2016} reported the detection of maser emission in two more SNRs, W44 and W51C, and in the candidate SNR G5.7-0.0. \citet{Li2017} carried out a 95.2~GHz CH$_3$OH emission survey towards eight SNRs, and methanol emission was detected in three SNRs near the Galactic Centre: Sgr~A East, G 0.1-0.1 and G 359.92-0.09.

The methanol molecule is formed on the surface of dust grains in molecular clouds, and there is no efficient route of methanol formation in the gas phase \citep{Garrod2006}. The sputtering of icy mantles of dust grains at the shock front leads to the ejection of methanol to the gas phase. The compressed and warm gas in the post-shock region is conducive to the collisional excitation of molecular species. The methanol molecules must survive the passage of the shock front -- the shock must be non-dissociative or C-type. The OH molecule is produced from H$_2$O via photo-dissociation and ion--molecule reactions in the post-shock region. The high ionization rate of the molecular gas is a necessary condition for the existence of OH masers \citep{Wardle1999}. Thus, the effectiveness of the maser pumping is a question both of the physical parameters in the interstellar gas (gas density, temperature, velocity gradient) and molecular abundance. As a rule, in the maser pumping modelling these parameters are independent variables and are varied in a wide range to find optimal conditions for the maser operation \citep[e.g.,][]{Nesterenok2016}. The shock model imposes constraints on the possible range of physical parameters in a maser source: the length of the C-type shock (in the direction of propagation) is inversely proportional to the pre-shock gas density and the ionization rate, while the number density of emitting molecules is regulated by chemical processes in the hot/warm shocked gas. Moreover, the gas temperature, velocity gradient and other parameters are not constant and vary with the coordinates. In our calculations we use the magnetohydrodynamic model of C-type shock coupled to a full gas--grain chemical network \citep{Nesterenok2018,Nesterenok2019}.

\citet{Flower2010b} reported the first calculations of the spectrum of methanol emerging from the shock waves in molecular outflows. Our study resembles that by \citet{Flower2010b} but differs in more extended chemical network used in our calculations. All important chemical processes that affect the methanol abundance in the shock wave are taken into account -- the production of methanol on the surface of dust grains, sputtering of icy mantles of dust grains at the shock front, various processes of gas-phase methanol destruction: adsorption on to dust grains, collisional dissociation, reactions with hydrogen atoms and ion-molecule reactions. We compute a large grid of shock models that vary in cosmic ray ionization rate, gas density and shock speed. Here, we emphasize the methanol and hydroxyl transitions that have population inversion, and infer the physical conditions necessary for the operation in one source of the hydroxyl maser 1720~MHz and Class I methanol masers simultaneously. In Section 2, we briefly describe the C-type shock model and our calculations of energy level populations of OH and CH$_3$OH molecules. Section 3 contains our results, and their discussion is presented in Section 4. In Section 5, we make our conclusions.

\section{Calculations}
\subsection{Chemistry}
The simulations of C-type shock start with the modelling of the chemical and thermal evolution of the static molecular gas. In these simulations, the species are assumed to be initially in atomic form except for hydrogen, which is assumed to be molecular. The initial elemental fractional abundances relative to the total H nucleus number density are given in Table~\ref{table:elabund} along with the choice of the ionization state \citep{Ruaud2016}. Following \citet{Walsh2010}, P- and F-containing species are eliminated in order to reduce computation time. The C/O elemental ratio in non-refractory material is equal to 0.7. We assume that the molecular cloud is shielded from the interstellar ultraviolet radiation field ($A_V = 10$), and the ionization rate $\zeta_\mathrm{H_2}$ of the molecular gas is suggested to be due to cosmic rays. Detailed information on the chemical model can be found in appendix~A of the paper by \citet{Nesterenok2018}.

The two-phase chemical model is used: the chemical reactions take place in the gas phase and on the surface of dust grains. The gas-phase chemical network is based on the UMIST database for astrochemistry (UDfA), 2012 edition \citep{McElroy2013}. Chemical species that are not included in the current edition of the UDfA network are added -- CH$_3$O, CH$_2$OH, HC$_2$O, CH$_3$CO, HOCO, CH$_3$OCH$_2$. Thus, our gas-phase network consists of reactions involving 430 species. Reactions of added species with abundant ions are taken into account \citep{Faure2010}. As additional chemical network updates, the gas-phase chemistry is supplemented by the list of neutral--neutral reactions published by \citet{Palau2017,Shingledecker2018}. Destruction rates of dimethyl ether and methyl formate by collisions with ions are taken from \citet{Ascenzi2019}. An update is done of the rate coefficients of the reactions involving carbon-chain species \citep{Chabot2013}. Photo-dissociation and photo-ionization rates of gas-phase species are updated according to \citet{Heays2017}. We have updated the rate of the reaction between CH$_3$OH and OH according to the theoretical study by \citet{Gao2018}. 

In total, 162 species are included in the grain surface chemical network. The desorption energies for chemical species are taken from \citet{Penteado2017}. The bimolecular reactions on the grain surface are taken from \textsc{nautilus} code files \citep{Ruaud2016}. The reaction activation barriers are also adopted from \textsc{nautilus} with one exception: the activation barrier of the reaction $\text{CO} + \text{O} \to \text{CO}_2$ is equal to 630~K \citep{Minissale2013}. Photo-dissociation and photo-ionization of chemical species adsorbed on the surface of dust grains are included \citep{Ruffle2001}. The solid/gas photo-dissociation coefficient ratio is taken to be equal to 0.15, where we take into account that a molecule located on the grain surface is partially shielded from incident radiation by the body of the grain \citep{Kalvans2018}. The photo-desorption of adsorbed species by interstellar and cosmic ray induced UV radiation is taken into account. We use a single photo-desorption yield $Y$ for all adsorbed species, $Y = 10^{-4}$ species per photon. For the CH$_3$OH molecule, it is assumed that photo-desorption process is dominated by the desorption of photo-fragments CO and H$_2$ \citep{Bertin2016}. The reactive, or chemical, desorption mechanism of adsorbed species is taken into account. For each surface reaction that leads to a single product, a proportion $f$ of the product species is released into the gas phase. In our previous calculations \citep{Nesterenok2018}, we used a simplified approach where the fraction $f$ was assumed to be fixed and equal to 0.01. Here we use more accurate approach based on the Rice-Ramsperger-Kessel theory \citep[see][and references therein]{Garrod2007}. The fraction $f$ of reactions resulting in desorption is:

\begin{equation}
\displaystyle
f = \frac{aP}{1 + aP}, \quad P = \left( 1 - \frac{E_\mathrm{D}}{E_\mathrm{reac}} \right)^{s-1},
\label{eq_reactive_des}
\end{equation}

\noindent
where $E_\text{D}$ is the desorption energy of the product molecule, $E_\text{reac}$ is the energy released in the reaction, $s$ is the number of vibrational modes in the molecule/surface-bond system. For diatomic species $s = 2$; for all others, $s = 3N-5$, where $N$ is the number of atoms in the molecule. The parameter $a$ is the ratio of the surface--molecule bond frequency to the frequency at which energy is lost to the grain surface \citep{Garrod2007}. Here we take $a = 0.01$. In the case $E_\text{D} << E_\text{reac}$, one has $P \approx 1$ and $f \approx a$ for small $a$. However, this condition is not satisfied for H-addition reactions in the chain of methanol production. For these reactions, the calculations based on equation (\ref{eq_reactive_des}) provide desorption factors $f$ that are a few times lower than the simple approach used in our previous work. The reactive desorption is the principal mechanism of injection of methanol molecules to the gas phase in cold gas in our chemical model. Another mechanism of methanol release may be direct cosmic ray induced desorption. However, the last mechanism is applied only for species with low desorption energy in our chemical model, $E_\text{D} \leq 1250$~K -- we use the approach by \citet{Roberts2007} to calculate the cosmic ray induced desorption rates.

Atomic hydrogen H is an important chemical agent that affects the abundances of many chemical species in the gas phase and on the surface of dust grains \citep[e.g.,][]{Padovani2018,Shaw2020}. In dark molecular clouds, atomic hydrogen is formed in (i) ion--molecule reactions induced by the propagation of cosmic rays in molecular gas and in (ii) direct dissociation by cosmic ray particles. The ion--molecule chemistry begins with the ionization of H$_2$ by cosmic ray particles. Hydrogen atoms are produced in the reactions

\begin{equation}
\begin{array}{l}
\displaystyle
\text{H}^+_2 + \text{H}_2 \to \text{H}^+_3 + \text{H}, \\ [5pt]
\displaystyle
\text{H}^+_3 + e^{-} \to \text{H} + \text{H} + \text{H}, \; \text{or} \; \mathrm{H_2} + \text{H},
\end{array}
\end{equation}

\noindent
and others. The direct dissociation of molecular hydrogen is entirely attributed to collisions with secondary electrons produced during the cosmic ray ionisation of molecular gas. \citet{Padovani2018} found that the ratio between cosmic ray dissociation and ionisation rates of molecular hydrogen is about 0.6--0.7. Note, this ratio is equal to 0.1 according to UDfA. The rate of direct H$_2$ dissociation is taken to be equal to 0.7 of the H$_2$ ionization rate provided by the UDfA chemical network.

A list of the collisional dissociation reactions taken into account is presented in appendix E in \citet{Nesterenok2018}. The reactions of collisional dissociation of methanol included in our chemical network are \citep{Baulch2005}:

\begin{equation}
\begin{array}{l}
\displaystyle
(1) \quad \mathrm{CH_3OH + M \to CH_3 + OH + M} \\ [10pt]
\displaystyle
(2) \quad \mathrm{CH_3OH + M \to CH_2OH + H + M}
\end{array}
\label{eq_meth_diss}
\end{equation}

\noindent
where M is a collisional partner. The rate coefficients provided by \citet{Baulch2005} are based on the studies of methanol pyrolysis in bath gas Ar. The most likely collisional partner in interstellar molecular gas is H$_2$, and we crudely rescale reaction rates multiplying by a factor of 3 \citep{Palau2017,Nesterenok2018}. The adopted parameters of the reactions (\ref{eq_meth_diss}) are $\alpha_1 = 2.5 \times 10^{-7}$, $\alpha_2 = 8.3 \times 10^{-8}$~cm$^3$~s$^{-1}$, $\beta_1 = \beta_2 = 0$, $\gamma_1 = \gamma_2 = 33080$~K. Collisional dissociation becomes the dominant channel of methanol destruction at neutral gas temperature $T_\text{g} \gtrsim 2000$~K. 

During the evolution of the molecular cloud, gas-phase molecules freeze out on dust grains. Methanol molecules are formed from CO by sequential hydrogenation steps on the surface of dust particles. The evolutionary time $t_0$ of the molecular cloud at which the methanol ice abundance relative to elemental hydrogen is equal to $10^{-5}$ is chosen as the starting point for the shock wave modelling. The methanol ice abundance considered is rather optimistic and corresponds to the upper limit of this parameter observed in molecular clouds \citep{Boogert2015}. We fix the methanol abundance at the start of shock wave simulations in order to remove the effect of the unknown evolutionary age of a molecular cloud. We focus on the post-shock processing of the methanol molecule at various parameters of the shock model, the initial methanol abundance being the same for all models.

\begin{table}
\centering
\caption{Initial elemental abundances with respect to H nucleus number density. \label{table:elabund}}
\begin{tabular}{llll}
\hline
Specimen & $n_\mathrm{i}/n_\mathrm{H,tot}$ & Specimen & $n_\mathrm{i}/n_\mathrm{H,tot}$ \\
\hline
H$_2$ & 0.5 & \quad Si$^{+}$ & $8 \times 10^{-9}$ \\
He & 0.09 & \quad Fe$^{+}$ & $3 \times 10^{-9}$ \\
N & $6.2 \times 10^{-5}$ & \quad Na$^{+}$ & $2 \times 10^{-9}$\\
O & $2.4 \times 10^{-4}$ & \quad Mg$^{+}$ & $7 \times 10^{-9}$ \\
C$^{+}$ & $1.7 \times 10^{-4}$ & \quad Cl$^{+}$ & $10^{-9}$ \\
S$^{+}$ & $8 \times 10^{-8}$ & & \\
\hline
\end{tabular}
\end{table}

\subsection{Parameters of the C-type shock model}
\citet{Crutcher2010} analysed samples of diffuse and molecular clouds with Zeeman observations in order to infer the distribution of the total magnetic field strength in the interstellar gas. They found that magnetic field strength in molecular clouds is randomly distributed between very small values and a maximum value that varies with the gas density:

\begin{equation}
B = \beta B_0 \left( n_\mathrm{H,tot}/n_0 \right)^{\alpha},
\label{eq_magn_field}
\end{equation}

\noindent
where $n_0 = 300$~cm$^{-3}$, $B_0 = 10 \umu G$, $\alpha = 0.65$, $0 < \beta \leq 1$, and $n_\text{H,tot} \geq n_0$ (see also \cite{Dudorov1991}). As an estimate of the magnetic field component in the pre-shock gas transverse to the gas flow direction, we use equation (\ref{eq_magn_field}) with the parameter $\beta = 1$ (maximal magnetic field strength). Note, we used the parametrization of the magnetic field strength by \citet{Crutcher1999} in our previous studies \citep{Nesterenok2019,Nesterenok2020}. The parametrizations by \citet{Crutcher1999} and \citet{Crutcher2010} give similar estimates of the magnetic field strength in molecular gas at $10^3 \lesssim n_\text{H,tot} \lesssim 10^5$~cm$^{-3}$. At $n_\text{H,tot} = 2 \times 10^6$~cm$^{-3}$, equation (\ref{eq_magn_field}) predicts a magnetic field twice as high as the approximation from earlier work.

Gas--grain interactions have an important effect on C-type shock waves propagating in molecular clouds \citep{Flower2003}. For gas densities $n_\mathrm{H_2} \gtrsim 10^4$~cm$^{-3}$, the grains dominate the momentum transfer between the neutral gas component and ions, and the heating of the neutral gas. Moreover, the properties of dust grains determine the rate of H$_2$ formation and, as a consequence, the abundance of atomic hydrogen in the cold molecular gas. In our model, the dust is assumed to be composed of spherical silicate particles of 0.05~$\umu$m in radius, and the dust--gas mass ratio is taken to be equal to 0.01. The specific surface area of dust is 10$^{-21}$~cm$^2$ per H. These dust parameters are reconciled with the mean H$_2$ production rate observed in  molecular clouds \citep{Nesterenok2019}.

In cold molecular clouds, the {\it ortho}-to-{\it para} ratio (OPR) of H$_2$ molecule slowly decays to a small value, $<< 1$ \citep{Lique2014}. {\it Ortho}-to-{\it para} conversion of the H$_2$ molecule is driven by proton exchange reactions with ions H$^+$ and H$_3^+$. The {\it ortho}-H$_2$ destruction is compensated by the formation of H$_2$ molecules on dust grains and by the destruction of ions. The time-scale of the {\it ortho}-to-{\it para} conversion can be significantly longer than the time-scale of dynamic evolution of the molecular gas at a low cosmic ray ionization rate \citep{Flower2006,Pagani2013}, and the H$_2$ OPR may not reach the steady state value. The observations of deuterated species and ammonia suggest that the H$_2$ OPR in molecular-cloud cores may lie in the range from values as low as $10^{-3}$ up to $\sim 1$ \citep[e.g.,][]{Pagani2009}. Here, the H$_2$ OPR is taken to be equal to 0.1 at the start of simulations of the chemical evolution of a molecular cloud – some arbitrary small value of the parameter. In a hot molecular gas behind interstellar shock, the reactive collisions (which include exchange of protons) of H$_2$ with H is the main mechanism of {\it ortho}-/{\it para}-H$_2$ interconversion. The {\it para}-to-{\it ortho}-H$_2$ conversion in the shock wave was considered in detail in our recent paper \citep{ Nesterenok2019}. In simulations, we take into account the following  processes of the {\it ortho}-/{\it para}-H$_2$ interconversion: reactive collisions H$_2$--H \citep{Lique2015,Bossion2018}, collisions H$_2$--H$^+$ \citep{Gonzalez-Lezana2017}, and H$_2$ formation on dust grains.

We have computed a grid of steady-state C-type shock models with shock speeds starting from 5~km~s$^{-1}$ up to the limiting value of the C-type shock speed. The profiles of the velocity, density, and temperature for each of the gas components (ions, electrons, and neutral particles) in the shock are calculated by integrating the mass, momentum, and energy conservation equations -- a detailed description of the governing equations is given in appendix~D in \citet{Nesterenok2018}. We consider the sputtering of icy mantles of dust grains by neutral species H$_2$, He and CO. The sputtering of dust grain cores and destruction of molecules in the sputtering process are not taken into account. The H$_2$ molecule is the main coolant of the hot/warm shocked gas, and the maximal speed at which C-type shock exists is regulated by collisional dissociation of the H$_2$ molecule. The rate of collisional dissociation of H$_2$ is calculated based on the distribution of energy level populations. The H$_2$ dissociation rate coefficients were published by \citet{Bossion2018} for H$_2$--H collisions, \citet{Martin1998,Ceballos2002} for H$_2$--H$_2$ collisions, and by \citet{Trevisan2002} for H$_2$--e$^{-}$ collisions. 

Table~\ref{table:modelparam} provides the shock parameters that are adopted in our calculations.

\begin{table}
\caption{Parameters of the shock model.}
\label{table:modelparam}
\begin{tabular}{p{3.5cm} p{3.5cm}}
\hline
Parameter & Value \\
\hline
Pre-shock gas density, $n_{\text{H,tot}}$ & $2 \times 10^4$--$2 \times 10^6$~cm$^{-3}$\\
Shock speed, $u_{\text{s}}$ & 5--70~km~s$^{-1}$ \\
Cosmic ray ionization rate, $\zeta_\mathrm{H_2}$ & $3 \times 10^{-17}$ -- $10^{-14}$~s$^{-1}$ \\
Initial {\it ortho}-to-{\it para}-H$_2$ ratio & 0.1 \\ 
Magnetic field strength, $\beta$ & 1 \\
Micro-turbulence speed, $u_{\text{turb}}$ & 0.3~km~s$^{-1}$ \\
Methanol ice abundance in the pre-shock gas & $10^{-5}$ \\
\hline
\end{tabular}

\medskip
The total number density of H nuclei is $n_\mathrm{H,tot} = n_\mathrm{H} + 2 n_\mathrm{H_2}$. For the definition of the parameter $\beta$ see equation (\ref{eq_magn_field}).
\end{table}

\subsection{Spectroscopic and collisional data}
The spectroscopic data on the OH molecule are taken from the HITRAN database \citep{Gordon2017} and those on the CH$_3$OH molecule are taken from \citet{Mekhtiev1999}.

The spin-orbit coupling of a single unpaired electron in an O atom splits the rotational energy levels of the OH molecule into $^2\Pi_{1/2}$ and $^2\Pi_{3/2}$ ladders. Each rotational level within these sets is further split into two close lying levels by the $\Lambda$-doubling. The two $\Lambda$-doublet levels have opposite total parities under a space-fixed inversion operator: levels with parity $(-1)^{j-1/2}$ are labelled $e$, and levels with parity $-(-1)^{j-1/2}$ are labelled $f$ \citep{Marinakis2019}. Each of the sublevels of the $\Lambda$-doublet is further split by the hyperfine interaction, and the resulting sublevels are characterized by different values of the total angular momentum quantum number $F$. We take into account 56 hyperfine resolved rotational energy levels of the OH molecule (the highest energy level is $^2\Pi_{1/2}$, $j = 6.5e$, $F = 7$ with energy of about 1550~K). The hyperfine resolved rate coefficients for collisions of OH with {\it ortho}- and {\it para}-H$_2$ were calculated by \citet{Offer1994}. These rate coefficients were initially calculated for 24 energy levels, but have since been extended to 48 energy levels \citep{Cragg2002}. The rate coefficients for collisions between OH and He atoms for the lowest 56 energy levels of the OH molecule were calculated by \citet{Marinakis2016,Marinakis2019}. The rate coefficients for OH--H$_2$ and OH--He collisions are available for gas temperatures up to 300~K. The rate coefficients for OH--H$_2$ collisions were kindly provided by Dr. Ostrovskii, and those for OH--He collisions -- by Dr. Marinakis. 

In the calculations, we consider energy levels of methanol with angular momentum quantum number $j \leq 15$ which belong to torsional states $v_t$ = 0, 1, 2. The total number of energy levels for each symmetry state of the molecule (A type and E type methanol) is 768. \citet{Rabli2010,Rabli2010b} calculated rate coefficients for rotational (torsionally elastic) transitions of methanol in collisions of methanol with He atoms and H$_2$ molecules. \citet{Rabli2011} provided rate coefficients for a limited set of torsionally inelastic transitions of methanol in collisions with He atoms. The collisional rate coefficients are available for gas temperatures $T_\mathrm{g} \leq 200$~K for torsionally elastic transitions, and for $T_\mathrm{g} \leq 400$~K for torsionally inelastic transitions. The energy level populations of A type and E type methanol are calculated independently. Here, the A and E symmetry species are assumed to be equally abundant.

The collisional coefficients are assumed to remain constant at gas temperatures above the maximal temperature for which collisional data are available.

\subsection{Calculation of molecular energy level populations}
The physical and chemical profiles that are derived from the shock wave modelling are used in calculations of energy level populations of CH$_3$OH and OH molecules. The shock profile is split into layers, and the energy level populations of species are calculated for each layer. The local statistical equilibrium is assumed in these calculations. This assumption may not be fulfilled at the shock peak, where physical parameters vary rapidly. However, in the post-shock region the flow time is sufficiently long that the assumption of local statistical equilibrium is justified \citep{Gusdorf2008,Flower2009}. 

The system of equations for energy level populations $n_{\text{i}}$ is

{\setlength{\mathindent}{0pt}
\begin{equation}
\begin{array}{l}
\displaystyle
\sum_{k=1, \, k \ne i}^M \left( R_{\mathrm{ki}} + C_{\mathrm{ki}} \right) n_{\mathrm{k}} - \\ [15pt]
\displaystyle
- n_{\mathrm{i}} \sum_{k=1, \, k \ne i}^M \left( R_{\mathrm{ik}} + C_{\mathrm{ik}} \right)=0, \quad i=1,...,M-1, \\ [15pt]
\displaystyle
\sum_{i=1}^M n_{\mathrm{i}} = 1,
\end{array}
\label{stat_eqn}
\end{equation}}

\noindent
where $M$ is the total number of energy levels, $R_\text{ik}$ are radiative rate coefficients, and $C_\text{ik}$ are collisional rate coefficients. The rate coefficients for downward and upward radiative transitions are

\begin{equation}
\begin{array}{l}
\displaystyle
R_\mathrm{ik}^{\downarrow} = B_\mathrm{ik} J_\mathrm{ik} + A_\mathrm{ik}, \\ [10pt]
\displaystyle
R_\mathrm{ki}^{\uparrow} = B_\mathrm{ki} J_\mathrm{ik},
\end{array}
\label{eq_radiat_trans}
\end{equation}

\noindent
where $A_\text{ik}$ and $B_\text{ik}$, $B_\text{ki}$ are the Einstein coefficients, $J_\text{ik}$ is the radiation intensity in spectral line averaged over the direction and over the line profile.

The large velocity gradient (LVG) or Sobolev approximation is used in the evaluation of the line intensities \citep{Sobolev1957}. In the cooling post-shock region, the gas temperature is much higher then the dust temperature and we disregard the dust emission in our calculations. The mean intensity in a molecular line (averaged over direction and line profile) can be calculated \citep{Hummer1985,Nesterenok2016}: 

\begin{equation}
\displaystyle
J(z) = S_\mathrm{L}(z) \left[1 - 2\mathcal{P}(\delta, \gamma) \right], 
\label{eq_line_intens1}
\end{equation}

\noindent
where $S_\mathrm{L}$ is the line source function, $\mathcal{P}(\delta, \gamma)$ is the one-sided loss probability function, $\delta$ and $\gamma$ are parameters equal to the gas velocity gradient normalized on the absorption coefficient in the continuum and molecular line, respectively. The emission of inverted transitions is not taken into account in the evaluation of energy level populations. 

The complexity of the OH energy level diagram leads to a set of radiative transitions in which certain transitions (connecting different rotational states) lie only a few MHz apart in frequency. Thus, emission in one line may be brought into resonance with the emission in another line through the thermal broadening and velocity gradient in the gas flow \citep{Litvak1969,Burdyuzha1973,Doel1990}. For each group of transitions connecting hyperfine splitting doublets of various rotational states of the OH molecule, the line overlap is considered for the pair of transitions having the closest frequencies. The mean intensity in one of the two overlapping lines can be calculated \citep{Nesterenok2020,Nesterenok2020b}:

\begin{equation}
\begin{array}{l}
\displaystyle
J_{1}(z) = S_\mathrm{L1}(z) \left[1 - 2\mathcal{P}_{11}(\delta, \gamma_{1}, \gamma_{2}, \Delta x) \right] \\ [10pt]
\displaystyle
+ S_\mathrm{L2}(z) \left[1 - 2\mathcal{P}_{12}(\delta, \gamma_{1}, \gamma_{2}, \Delta x) \right],
\end{array}
\label{eq_line_intens2}
\end{equation}

\noindent
where parameters with the subscripts 1 and 2 refer to the first and second line, respectively, $\Delta x = (\nu_{1} - \nu_{2})/\Delta \nu_\mathrm{D}$ is the difference in the line frequencies normalized on the line profile width. The functions $\mathcal{P}_{11}$ and $\mathcal{P}_{12}$ were calculated by \citet{Nesterenok2020b}. The simultaneous overlap of three and more lines is not considered. Line overlap in methanol is expected to be unimportant \citep{Cragg2002,McEwen2014}.
 
The criterion of the applicability of the Sobolev approximation is the restriction on the length scale of the change in physical parameters $l$: 

\begin{equation}
\displaystyle
l \gg \Delta z_\mathrm{S}, \quad \Delta z_\mathrm{S} = u_\mathrm{D} \left\vert \frac{\mathrm{d}u}{\mathrm{d}z} \right\vert^{-1},
\label{eq_sobolev_length}
\end{equation}

\noindent
where $\Delta z_\mathrm{S}$ is the size of the region, where the radiation at a given frequency interacts with the medium -- the Sobolev length, $u(z)$ is the gas velocity, $u_\mathrm{D}$ is the line profile width in velocity units. As an estimate of $l$ we may take \citep{Gusdorf2008}:

\begin{equation}
\displaystyle
l \sim T_\mathrm{g}/(\mathrm{d}T_\mathrm{g}/\mathrm{d}z)
\end{equation}

\noindent
In the cooling post-shock gas $l \approx (2-5) \Delta z_\mathrm{S}$. For a pair of lines with frequency difference $\vert\Delta x\vert > l/\Delta z_\mathrm{S}$, equations (\ref{eq_line_intens2}) are no longer valid, and it makes no sense to take into account the line overlap using this formalism \citep{Nesterenok2020,Nesterenok2020b}. Let us define $N_\text{S}$ as the column density of molecules along the distance $\Delta z_\text{S}$ in the direction of shock propagation. It is this parameter that determines the radiative transfer in molecular lines in the medium, not the full column density of molecules along the gas flow.

The system of statistical equilibrium equations (\ref{stat_eqn}) is non-linear. The solution is obtained iteratively, and calculations stop when the maximum relative change in energy level populations is $<10^{-5}$ for two successive iterations. An acceleration of the iterative series may be achieved by applying the convergence optimization method \citep{Ng1974}.

\subsection{Maser parameters}
The expression for the gain of a molecular line $i \to k$ in the one-dimensional model of a flat gas--dust cloud is

\begin{equation}
\displaystyle
\gamma_\mathrm{ik}(z,\mu, \nu)=\frac{\lambda^2 }{8 \pi} A_\mathrm{ik} n_\mathrm{mol} \left(n_\mathrm{i}-\frac{g_\mathrm{i}}{g_\mathrm{k}} n_\mathrm{k} \right) \phi(z,\mu, \nu) - \kappa_\mathrm{c},
\label{eq_gain}
\end{equation}

\noindent
where $\mu$ is the $\text{cos}$ of the angle $\theta$ between the gas flow velocity and the line of sight direction, $n_\text{mol}$ is the molecule concentration (hydroxyl, A- or E-symmetry species of methanol), $g_\text{i}$ and $g_\text{k}$ are statistical weights of energy levels, $\kappa_\text{c}$ is the absorption coefficient of the dust. The spectral profile of the emission and absorption coefficients in the laboratory frame of reference is 

\begin{equation}
\phi(z,\mu, \nu) = \tilde{\phi}_\mathrm{ik} \left[ \nu - \nu_\mathrm{ik} \mu u(z) / c \right]
\label{eq_profile_lab}
\end{equation}

\noindent
where $\nu_\mathrm{ik}$ is the transition frequency, $\tilde{\phi}_\mathrm{ik}(\nu)$ is the normalized spectral line profile in the co-moving frame of the gas. The optical depth in the given maser transition in the direction $\mu$ to the shock propagation is

\begin{equation}
\displaystyle
\vert \tau_{\mu}(\nu) \vert = \frac{1}{\mu}\int \, \mathrm{d}z \, \gamma_\mathrm{ik}(z,\mu, \nu),
\label{eq_tau}
\end{equation}

\noindent
where the integration is over the region with positive gain, $\gamma_\mathrm{ik} > 0$. Here we call the parameter $1/\mu$ the aspect ratio -- the ratio of the amplification path along the line of sight to the shock width. In reality, the maximal value of the aspect ratio is limited by the dimensions of the shock front. Usually, a characteristic value of $\sim 10$ is adopted for the aspect ratio of strong masers in theoretical models \citep[e.g.,][]{Elitzur1989}. 

The maser action reduces the population inversion of energy levels. The production rate of maser photons approaches a limiting value as the intensity of maser radiation increases -- the maser becomes saturated. The volume emissivity of a fully saturated maser in the transition $i \to k$ is \citep{Neufeld1991}:

\begin{equation}
\displaystyle
\Phi_\mathrm{sat} = \frac{n_\text{i}/g_\text{i} - n_\text{k}/g_\text{k}}{(\Gamma_\text{i} g_\text{i})^{-1} + (\Gamma_\text{k} g_\text{k})^{-1}} n_\text{mol},
\label{eq_photon_prodrate}
\end{equation} 

\noindent
where energy level populations are calculated in the absence of saturated maser action, $\Gamma_\text{i}$ and $\Gamma_\text{k}$ are decay rates of upper and lower levels, respectively. The parameter $\Gamma_\text{i}$ takes into account collisional and radiative processes other than emission in the masing transition:

\begin{equation}
\displaystyle
\Gamma_\text{i} = \sum_{j=1, \, j \ne i,k}^M  R_{\mathrm{ij}} + \sum_{j=1, \, j \ne i}^M C_{\mathrm{ij}},
\label{eq_loss_rate}
\end{equation}

\noindent
The radiation intensity and the optical depth at which the maser becomes saturated can be estimated from the equality condition between the stimulated emission rate and the decay rate of energy levels. 

\citet{Elitzur1989} introduced the maser emission measure, $\xi$, which characterizes the efficiency of maser pumping,

\begin{equation}
\xi = \left(\frac{n_\mathrm{mol}}{\mathrm{1 \,cm^{-3}}}\right) \left(\frac{n_\mathrm{H_2}}{10^6 \,\mathrm{cm}^{-3}}\right) \left(\frac{1 \,\mathrm{km \, s^{-1} \, pc^{-1}}}{\vert \mathrm{d}u/\mathrm{d}z \vert}\right),
\label{eq_emiss_meas}
\end{equation}

\noindent
where normalization of the parameter $\xi$ is the same as in \citet{Leurini2016} \footnote{$1 \,\mathrm{km \, s^{-1} \, pc^{-1}} = 3.24 \times 10^{-14}$~cm~s$^{-1}$~cm$^{-1}$.}.

\begin{figure}
\includegraphics[width=85mm]{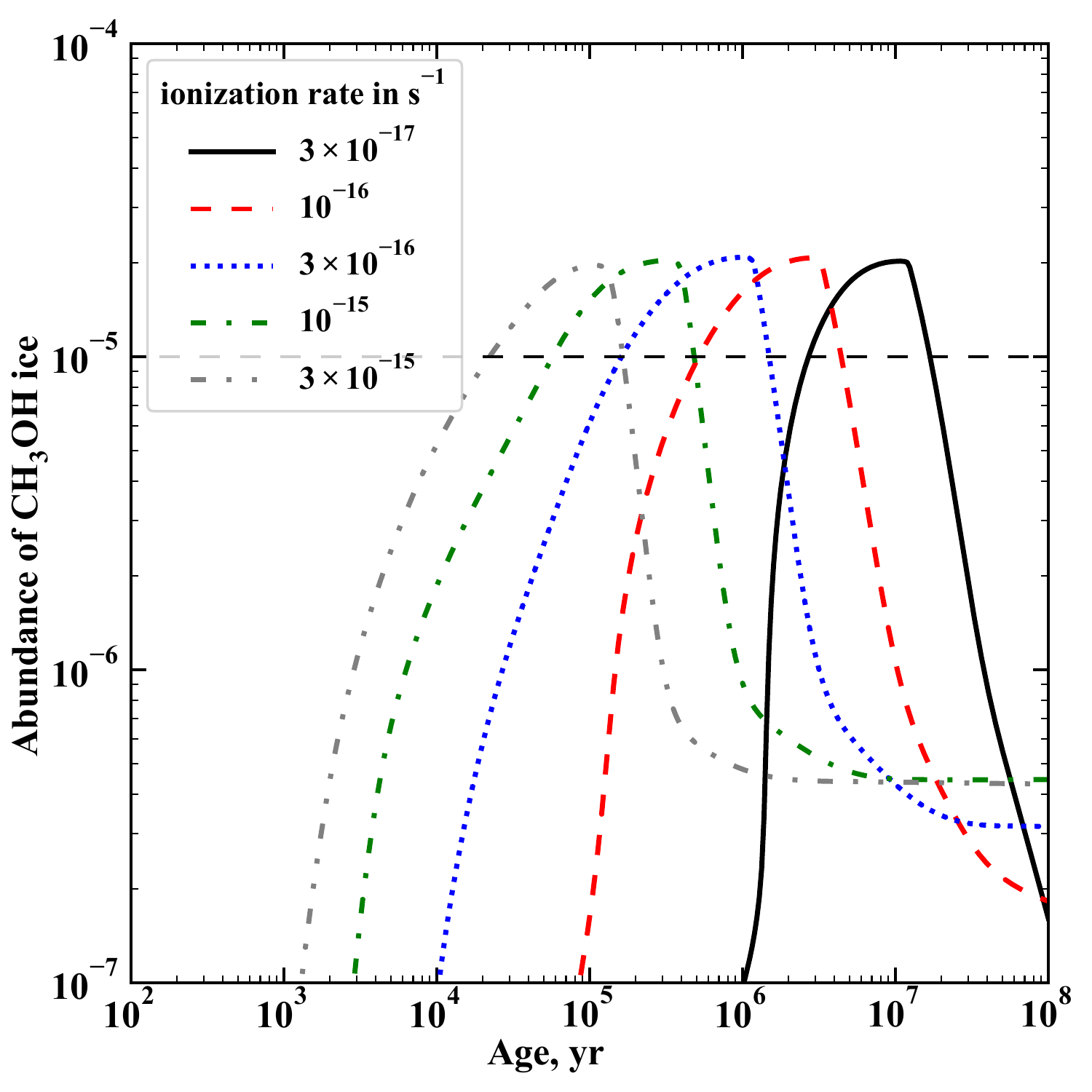}
\caption{The abundance of methanol ice relative to the total H nuclei number density as a function of the evolutionary age of the molecular gas. The results are presented for gas density $n_\text{H,tot} = 2 \times 10^5$~cm$^{-3}$ and for different values of cosmic ray ionization rate.}
\label{fig1}
\end{figure}

\section{Results}
\subsection{Chemical evolution of the dark cloud}
Fig.~\ref{fig1} shows the abundance of methanol adsorbed on dust grains as a function of the evolutionary age of the molecular gas. The methanol ice abundance grows as the evolutionary time of the cold molecular gas increases, and reaches a maximum of about $2 \times 10^{-5}$. At late times, the methanol converts into solid methane, and eventually most of the carbon ends up in the form of a mixture of surface-bound hydrocarbon molecules, see also \citet{Garrod2007}. The ion concentration and the gas temperature are higher when the gas ionization rate is higher, and the chemical evolution proceeds faster. The evolutionary times $t_0$ at which methanol ice abundance is equal to $10^{-5}$ are $2.7 \times 10^6$ and $2.2 \times 10^4$~yr at cosmic ray ionization rates $3 \times 10^{-17}$ and $3 \times 10^{-15}$~s$^{-1}$, respectively (at gas density $n_\text{H,tot} = 2 \times 10^5$~cm$^{-3}$).

The higher the cosmic ray ionization rate, the higher the abundance of atomic hydrogen in the gas phase that triggers the formation of hydrocarbon and carbohydrate molecules including methanol on the grain surface, see also \citet{Liu2020}. On the other hand, the main destruction channel of methanol adsorbed on dust grains is photo-dissociation by cosmic ray-induced UV, the reaction products being CH$_3$ and OH. The photo-dissociation of adsorbed methanol molecules is much less efficient than methanol destruction in the gas phase (in ion--molecule and photo-dissociation reactions). As a result, the methanol formation on the surface of dust grains is efficient even at the highest cosmic ray ionization rate in question. In our model, the reactive desorption mechanism is the principal mechanism of methanol ejection to the gas phase in cold molecular gas (the photo-desorption of methanol by cosmic ray-induced UV radiation is accompanied by methanol dissociation, and direct cosmic ray desorption is not used for species with high desorption energy). 

The abundance of H$_2$O ice lies in the range $10^{-5}$--$10^{-4}$ at the moment when the methanol production on the dust grain surface becomes efficient. At the start of shock wave simulations, the abundance of methanol ice relative to oxygen in the form of H$_2$O ice or atomic oxygen in the gas phase is about 0.1. In the shock, the gas-phase atomic oxygen turns into H$_2$O in chemical reactions. According to observational data, the abundance of solid-state H$_2$O in quiescent dark cloud cores lies in the range $10^{-5}$--$10^{-4}$, and methanol ice abundance with respect to water ice is 5--10 per cent \citep{Boogert2015,Goto2021}.

\begin{figure*}
\includegraphics[width=170mm]{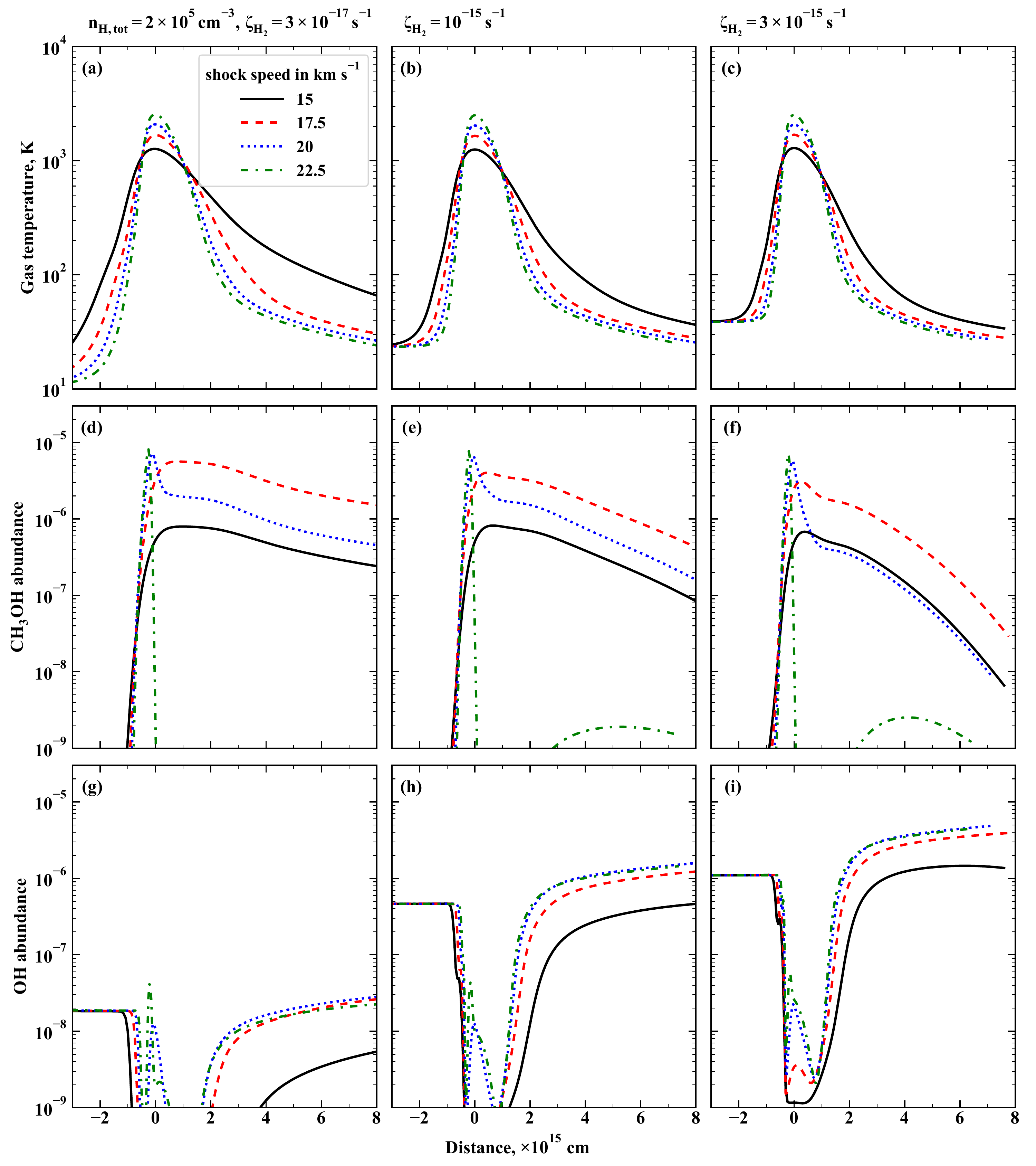}
\caption{Gas temperature, abundance of methanol and hydroxyl molecules as a function of distance in the C-type shock wave. The zero point in the horizontal axis is the point of the maximum temperature of the gas. The pre-shock gas density is $n_{\rm{H,tot}} = 2 \times 10^5$~cm$^{-3}$. The plots in each column correspond to models with cosmic ray ionization rates $3 \times 10^{-17}$, 10$^{-15}$, and $3 \times 10^{-15}$~s$^{-1}$. The results for shock speeds 15, 17.5, 20 and 22.5~km~s$^{-1}$ are presented on each plot.}
\label{fig2}
\end{figure*}

\subsection{Abundance of CH$_3$OH and OH molecules in the shock}
Fig.~\ref{fig2} shows the abundance of methanol and hydroxyl in the C-type shock, the pre-shock gas density is $n_\mathrm{H,tot} = 2 \times 10^5$~cm$^{-3}$. The results in the left-hand panels correspond to the cosmic ray ionization rate $\zeta_\mathrm{H_2} = 3 \times 10^{-17}$~s$^{-1}$ -- a typical value for dense molecular clouds \citep{Dalgarno2006}. The middle and right-hand panels in Fig.~\ref{fig2} correspond to cosmic ray ionization rates $10^{-15}$ and $3 \times 10^{-15}$~s$^{-1}$, respectively -- appropriate values for molecular clouds located close to SNRs \citep{Vaupre2014,Shingledecker2016}. At shock speed $u_\text{s} = 15$~km~s$^{-1}$, the sputtering of icy mantles of dust grains is not complete, and not all methanol adsorbed on dust grains is released to the gas phase, see Fig.~\ref{fig2} (d)--(f). At shock speed $u_\text{s} = 20$~km~s$^{-1}$, the sputtering of icy mantles of dust grains is complete. However, methanol molecules ejected to the gas phase are partially destroyed via reactions with H atoms and collisional dissociation reactions. The gas-phase methanol is almost entirely destroyed at the shock front at shock speeds $u_\text{s} \geq 22.5$~km~s$^{-1}$, the main destruction channel being collisional dissociation. The methanol is restored in the gas phase downstream  via reactive desorption mechanism. The methanol abundance in the gas phase is of the order of $10^{-9}$--$10^{-8}$ in the post-shock region at high shock speeds and at high cosmic ray ionization rates. The gas-phase methanol abundance in the post-shock gas is higher than its abundance in the pre-shock gas, see also \citet{Nesterenok2018}. 
% observations? Kristensen 2010, Sokolova 2020...

The methanol abundance in the gas phase just behind the shock peak is relatively high, about $10^{-6}$--$10^{-5}$, at $15 \leq u_\text{s} \leq 20$~km~s$^{-1}$ and $n_\mathrm{H,tot} = 2 \times 10^5$~cm$^{-3}$, see Fig.~\ref{fig2} (d)--(f). It is likely that a fraction of molecules could be destroyed in the sputtering process \citep{Suutarinen2014} -- we do not consider this effect in our model. At low cosmic ray ionization rates, the main destruction channel of gas-phase methanol in the post-shock region is adsorption on to dust grains (the dust temperature in the shock is lower than the evaporation limit for methanol $T_\text{d,0} \approx 70$~K). The time-scale of this process is 

\begin{equation}
t \sim 10^3 \left( \frac{10^6 \, \mathrm{cm}^{-3}}{n_\mathrm{H_2}} \right) \left(\frac{50 \, \mathrm{K}}{T_\mathrm{g}}\right)^{1/2} \, \mathrm{yr}.
\end{equation}

\noindent
At high cosmic ray ionization rates, $\zeta_\mathrm{H_2} \gtrsim 10^{-15}$~s$^{-1}$, the ion--molecule reactions and photo-dissociation by cosmic ray-induced UV radiation become important in the post-shock evolution of methanol abundance. The abundance of methanol is rapidly declined in the post-shock region at high cosmic ray ionization rates, see Fig.~\ref{fig2} (f). The time of the gas flow from shock maximum up to the point where the gas temperature falls below 30~K is about $700-800$~yr at pre-shock gas density $n_\text{H,tot} = 2 \times 10^5$~cm$^{-3}$ and shock speed $u_\text{s} = 17.5$~km~s$^{-1}$. At the moment when the gas temperature constitutes 50~K, the time-scales of methanol destruction are about 600 and 200~yr at $\zeta_\mathrm{H_2} = 3 \times 10^{-17}$ and $\zeta_\mathrm{H_2} = 3 \times 10^{-15}$~s$^{-1}$, respectively. The contribution of the reaction between methanol and hydroxyl to the total rate of methanol destruction is small. 

The injection of CH$_3$OH and H$_2$O from grains into the gas phase occurs simultaneously. H$_2$O is a precursor of OH. In hot gas at the shock front, the equilibrium of the reversible reaction $\text{H}_2 + \text{OH} \rightleftarrows \mathrm{H_2O} + \text{H}$ is shifted to H$_2$O formation. In the cooling gas downstream, the OH production channels are photo-dissociation of the H$_2$O molecule and dissociative recombination reactions involving the H$_3$O$^+$ ion \citep{Wardle1999,Nesterenok2020}. The contribution of gas-phase methanol destruction to OH production is small, $<10$ per cent. The OH abundance increases in the cooling post-shock gas up to some equilibrium value, where OH production is compensated by OH destruction: the adsorption on to dust grains, reactions with atoms O, H, N and etc., see Fig.~\ref{fig2} (g)--(i). The higher the cosmic ray ionization rate, the higher the number density of OH molecules. The conditions that are favourable for effective OH production (high ionization rate) are not favourable for methanol conservation in the gas phase.

\begin{figure}
\includegraphics[width=85mm]{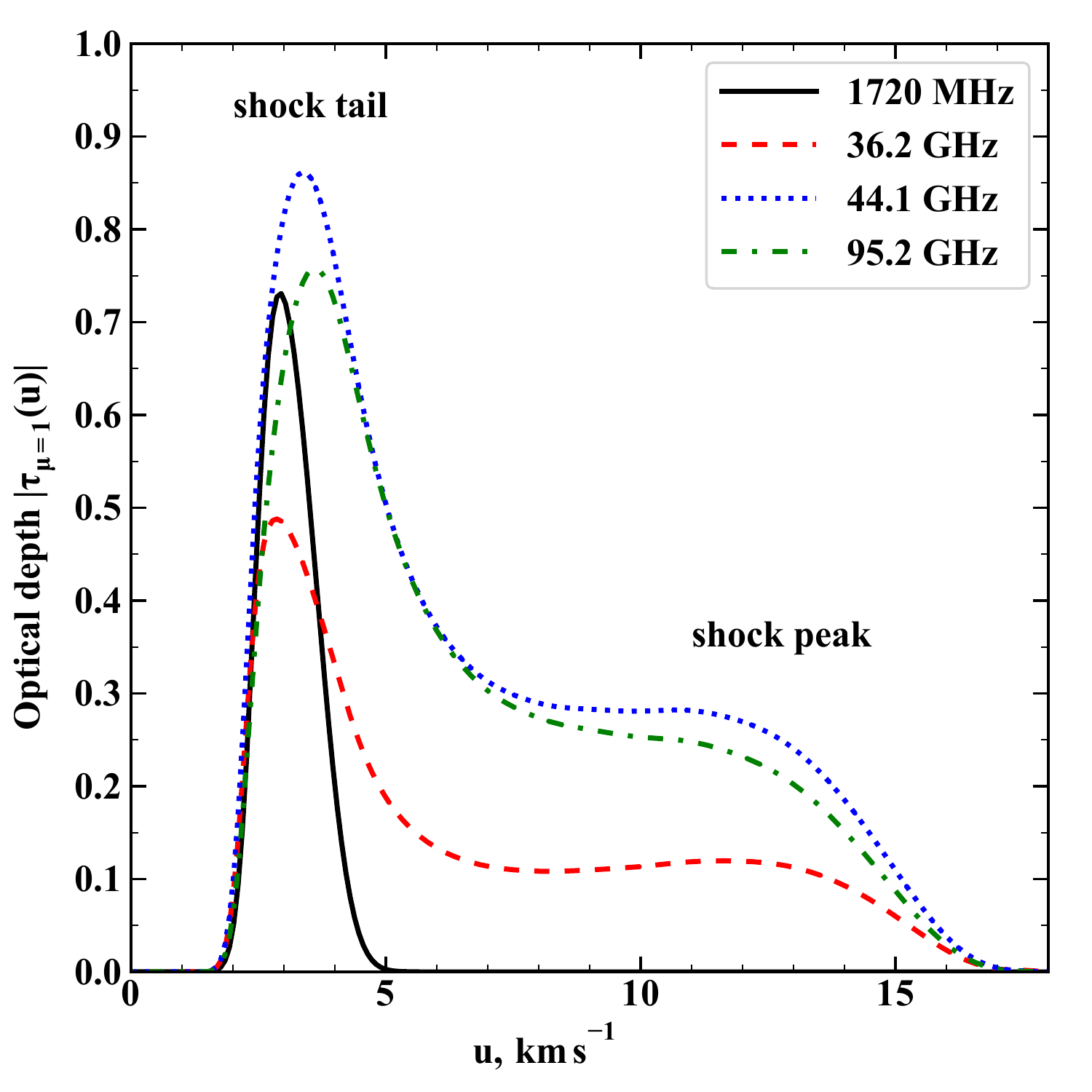}
\caption{The dependence of optical depth $\vert \tau_{\mu=1} \vert$ in maser transitions of hydroxyl (1720~MHz) and methanol (36.2, 44.1, 95.2~GHz) as a function of frequency shift in velocity units. The reference frame where the shock front is at rest is adopted, and the gas flow is moving away from the observer. The pre-shock gas has line-of-sight velocity $u_\text{s}$, and the far post-shock region has velocity $1-2$~km~s$^{-1}$. The shock model parameters are: $n_\text{H,tot} = 2 \times 10^5$~cm$^{-3}$, $\zeta_\mathrm{H_2} = 3 \times 10^{-15}$~s$^{-1}$, $u_\text{s}  = 17.5$~km~s$^{-1}$.}
\label{fig3}
\end{figure}

\begin{figure*}
\includegraphics[width=170mm]{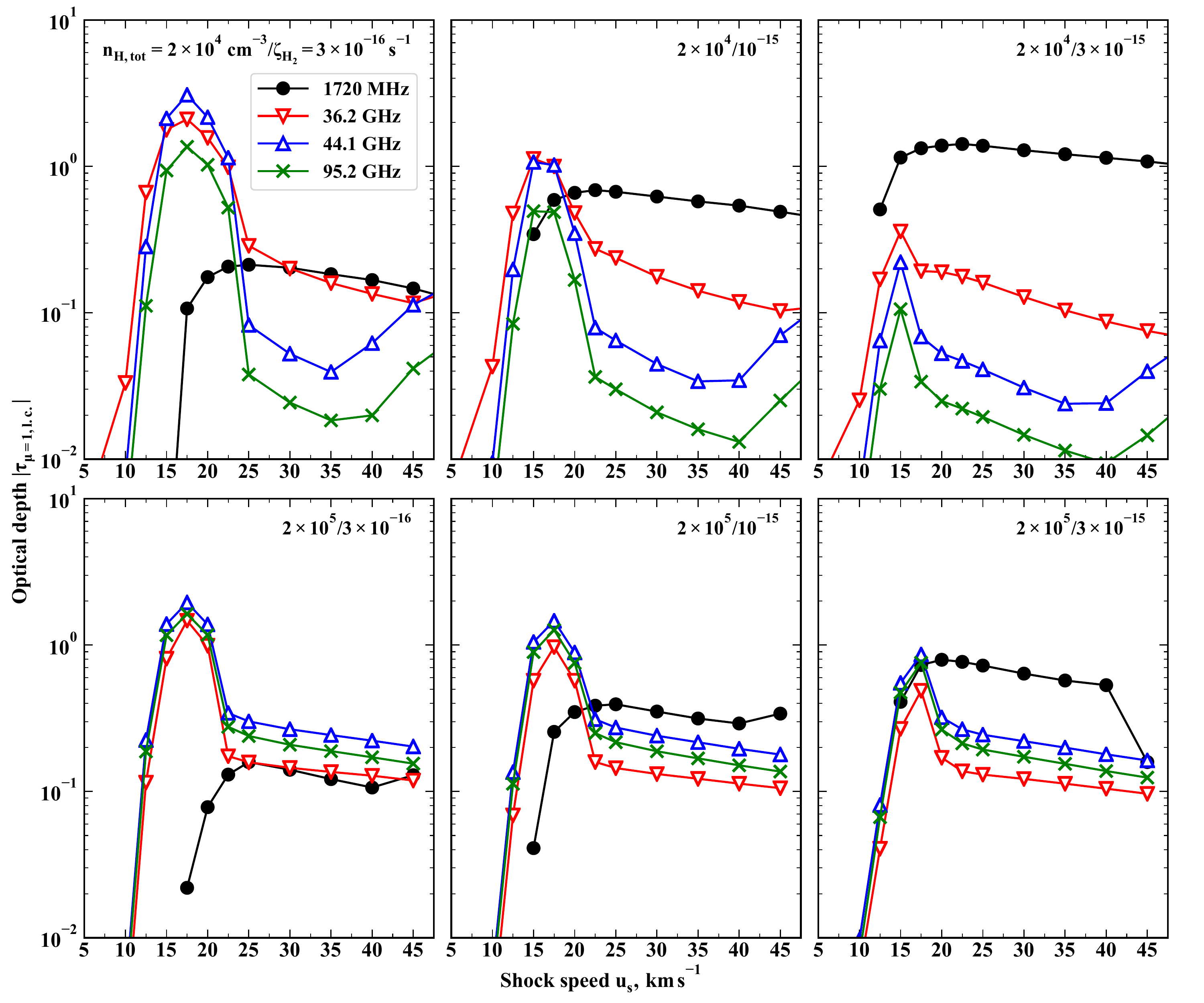}
\caption{Optical depth along the gas flow direction and at the line centre $\vert \tau_{\mu=\text{1,\,l.c.}} \vert$ in the OH (1720~MHz) and CH$_3$OH (36.2, 44.1, and 95.2~GHz) maser transitions. Each point corresponds to a particular shock model, the shock speed is along the horizontal axis. The values of parameters $n_\mathrm{H,tot}$ and $\zeta_\mathrm{H_2}$ are indicated on each graph.}
\label{fig4}
\end{figure*}

\subsection{Optical depth in CH$_3$OH and OH maser transitions}
Fig.~\ref{fig3} shows the dependence of optical depth $\vert \tau_{\mu=1} \vert$ in maser transitions along the gas flow direction as a function of frequency shift in velocity units. The results are presented for CH$_3$OH maser transitions at 36.2, 44.1, and 95.2~GHz and for the OH maser transition at 1720~MHz. The shock parameters are such that both hydroxyl and methanol masers exist: $n_\text{H,tot} = 2 \times 10^5$~cm$^{-3}$, $\zeta_\mathrm{H_2} = 3 \times 10^{-15}$~s$^{-1}$ and $u_\text{s}  = 17.5$~km~s$^{-1}$. Hydroxyl emission originates in the shock tail as the hydroxyl is formed in the cooling post-shock gas, while the zone of methanol masers starts at the shock peak and extends to the entire shock tail. The velocity gradient decreases in the shock tail and, as a result, the Sobolev length $\Delta z_\text{S}$ and column density of methanol molecules $N_\text{S}$ increase. The post-shock region with $u \lesssim 5$~km~s$^{-1}$ makes a major contribution to the optical depth along the gas flow. The gas temperature in this region is $T_\text{g} \lesssim 500$~K. Analogous results for methanol emission were obtained by \citet{Flower2010b}. 

The peaks of the optical depth in the hydroxyl transition at 1720~MHz and methanol transition at 36.2~GHz are close in frequency (see Fig.~\ref{fig3}), while the velocity shift between optical depth peaks in the hydroxyl transition and methanol transition at 44.1~GHz is about 0.4~km~s$^{-1}$ (for the line of sight along the outflow direction). The velocity shift between optical depth peaks in methanol transitions 44.1 and 95.2~GHz is about 0.25~km~s$^{-1}$ at model parameters in question. The optical depth $\vert \tau_\mu \vert$ along the line of sight at an angle $\theta$ to the direction of the gas flow can be crudely estimated as $\vert \tau_{\mu=1} \vert/\mu$, $\mu = \text{cos} \theta$. However, in the case of methanol, $\vert \tau_\mu \vert$ increases with decrease of $\mu$ substantially faster than $\vert \tau_{\mu=1} \vert/\mu$ as the radiation coming from different parts of the shock region is in resonance at high aspect ratio $1/\mu$, see equation (\ref{eq_profile_lab}). The velocity shift between emission lines is proportional to $\mu$. 

Fig.~\ref{fig4} presents the calculations of the optical depth at the line centre along the direction of shock propagation $\vert \tau_{\mu=\text{1,\,l.c.}} \vert$ in hydroxyl and methanol maser transitions as a function of shock speed $u_\text{s}$. The calculations are presented for cosmic ray ionization rates $3 \times 10^{-16}$, $10^{-15}$ and $3 \times 10^{-15}$~s$^{-1}$, and pre-shock gas densities $n_\text{H,tot} = 2 \times 10^4$ and $2 \times 10^5$~cm$^{-3}$. The release of H$_2$O and CH$_3$OH molecules locked in icy mantles of dust grains takes place at shock speeds $u_\mathrm{s} \gtrsim 15$~km~s$^{-1}$. This limits the shock speeds required for efficient maser operation from below. At high shock speeds, the absolute value of the optical depth in methanol lines is low due to methanol destruction at the shock front. In this case, the main contribution to the methanol emission is provided by the hot gas at the shock front. The efficient operation of methanol masers is possible in the narrow range of shock speeds 15--22.5~km~s$^{-1}$. The optical depth in methanol maser transitions is maximal at low cosmic ray ionization rates. The higher the cosmic ray ionization rate is, the higher the methanol destruction rate becomes, and the lower the absolute value of the optical depth in methanol maser transitions. While a high cosmic ray ionization rate is a necessary condition for OH molecule production in the post-shock region. Even at an enhanced cosmic ray ionization rate $\zeta_\mathrm{H_2} = 3 \times 10^{-16}$~s$^{-1}$ (left-hand panels in Fig.~\ref{fig4}), the absolute value of the optical depth in the OH transition at 1720~MHz is low, $\vert \tau_{\mu=\text{1,\,l.c.}} \vert \lesssim 0.2$. Nevertheless, the operation of both hydroxyl and methanol masers is possible at high cosmic ray ionization rates $\zeta_\mathrm{H_2} = 10^{-15}$--$3 \times 10^{-15}$~s$^{-1}$ and a narrow range of shock speeds $u_\text{s} \approx$ 15--20~km~s$^{-1}$. The absolute value of the optical depth in maser transitions is $\vert \tau_{\mu=\text{1,\,l.c.}} \vert \lesssim 1$ in this case. 

At high cosmic ray ionization rate and low gas density, the methanol maser transition 36.2~GHz has the highest optical depth in absolute value (the middle and right-hand panels in the upper row in Fig.~\ref{fig4}). Even at high shock speeds, optical depth in the 36.2~GHz transition is $\vert \tau_{\mu=\text{1,\,l.c.}} \vert \approx$ 0.1--0.2. At high gas density, the maser transition 36.2~GHz has the lowest optical depth among methanol transitions shown in Fig.~\ref{fig4}. At cosmic ray ionization rate $\zeta_\mathrm{H_2} \lesssim 10^{-16}$~s$^{-1}$ and shock speed $u_\text{s} \lesssim$ 20--25~km~s$^{-1}$, the {\it para}- to {\it ortho}-H$_2$ conversion in the shock is inefficient. In this case, the main collisional partner of molecules is {\it para}-H$_2$ (or {\it para}-H$_2$ and {\it ortho}-H$_2$ are equally important). At these physical conditions, the population inversion in the OH maser transition 1720~MHz is suppressed \citep{Pavlakis1996b,Nesterenok2020}. Exactly at these shock parameters, the absolute value of the optical depth in methanol transitions is the highest. However, the H$_2$ OPR does not seem to be a crucial factor in methanol maser pumping.

\subsection{Methanol maser transitions}
\begin{figure*}
\includegraphics[width=170mm]{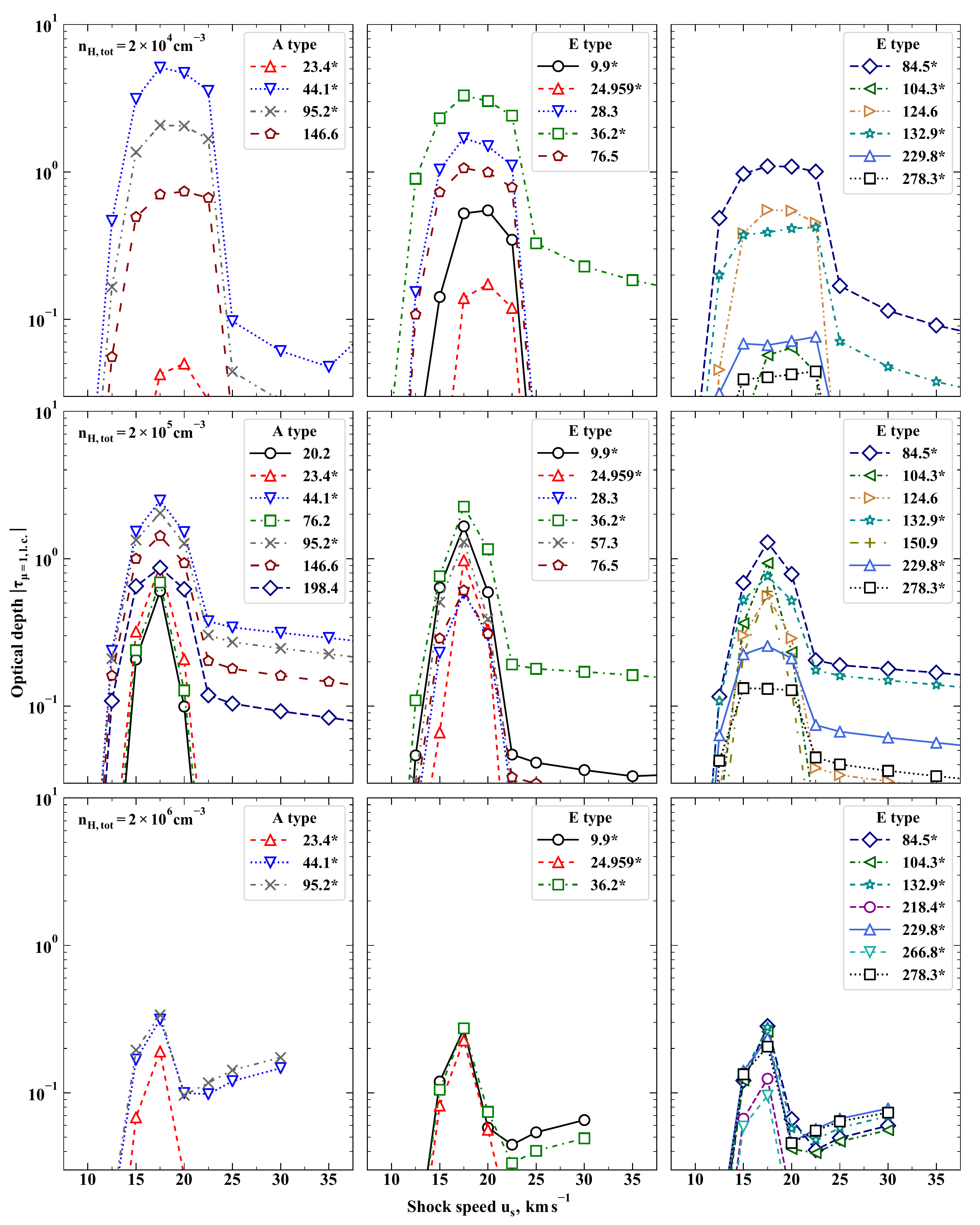}
\caption{Optical depth $\vert \tau_{\mu=\text{1,\,l.c.}} \vert$ for methanol maser transitions. The detected masers are marked by an asterisk \citep{Ladeyschikov2019}. The cosmic ray ionization rate is $\zeta_\mathrm{H_2} = 3 \times 10^{-17}$~s$^{-1}$, and the pre-shock gas density is $n_\mathrm{H,tot}= 2 \times 10^4$~cm$^{-3}$ (top), $n_\mathrm{H,tot}= 2 \times 10^5$~cm$^{-3}$ (middle), and $2 \times 10^6$~cm$^{-3}$ (bottom). Only one delegate of the transition series E~$J_2 \to J_1$ with $J = 5$ at 24.959~GHz is shown.}
\label{fig5}
\end{figure*}

\begin{figure*}
\includegraphics[width=170mm]{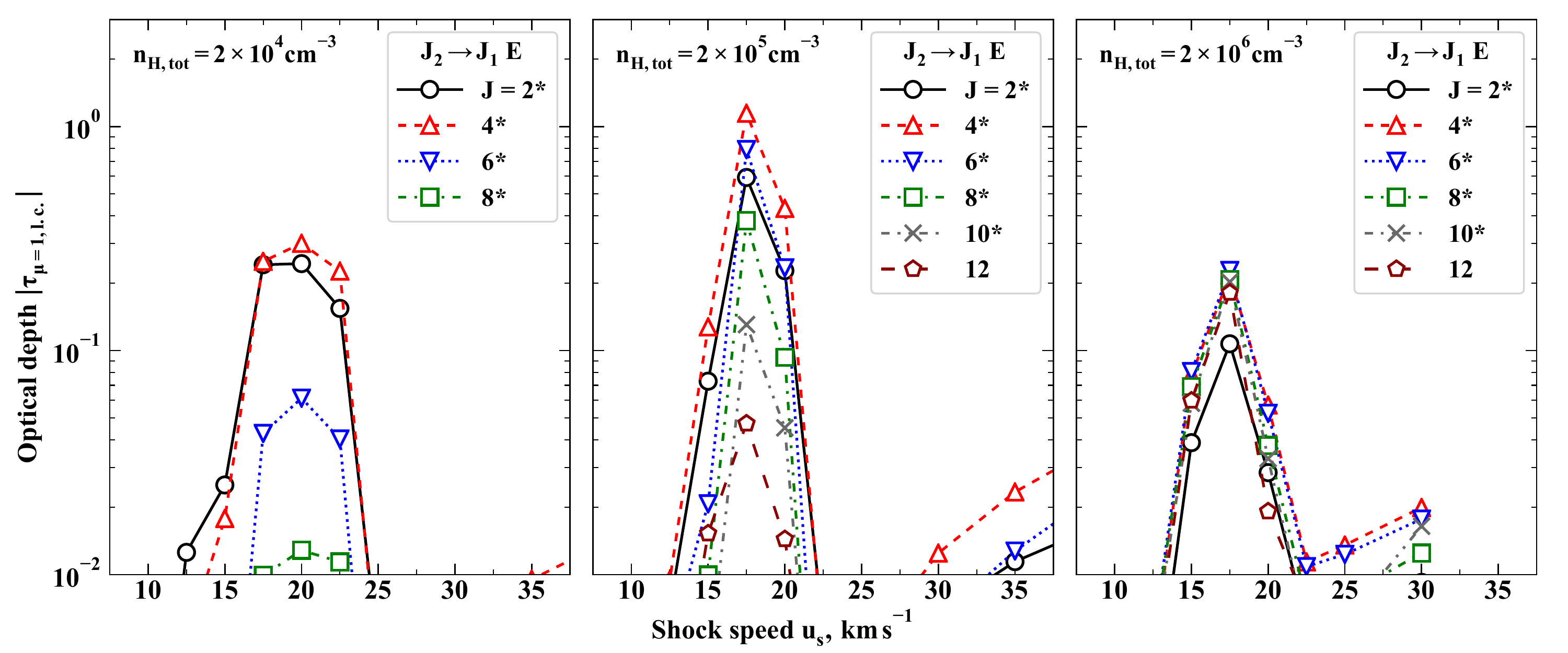}
\caption{Optical depth $\vert \tau_{\mu=\text{1,\,l.c.}} \vert$ for the series of E~type methanol transitions $J_2 \to J_1$. The detected masers are marked by an asterisk \citep{Voronkov2006,Towner2017}. The cosmic ray ionization rate is $\zeta_\mathrm{H_2} = 3 \times 10^{-17}$~s$^{-1}$ and the pre-shock gas density is indicated on each panel.}
\label{fig6}
\end{figure*}

Figs.~\ref{fig5} and \ref{fig6} show the optical depth for a large sample of methanol transitions. The results are shown for cosmic ray ionization rate $3 \times 10^{-17}$~s$^{-1}$, and pre-shock gas densities $n_\mathrm{H,tot} = 2 \times 10^4$, $2 \times 10^5$, and $2 \times 10^6$~cm$^{-3}$. All known detected Class I methanol maser transitions are included in this sample -- these transitions are marked by an asterisk \citep{Ladeyschikov2019}. The non-detected lines that have optical depth $\vert \tau_{\mu=\text{1,\,l.c.}} \vert > 0.5$ in at least one of the shock models are shown in Fig.~\ref{fig5}. At high aspect ratio, these transitions are expected to be strong masers according to our simulations. The methanol maser transitions are combined into series: the widespread A type methanol transitions 44.1 and 95.2~GHz and non-detected transitions 146.6 and 198.4~GHz belong to the series $J_0 \to (J-1)_1$. The numerous series of E type methanol masers is $J_{-1} \to (J-1)_0$ -- the detected transitions 36.2, 84.5, 132.9, 229.8 and 278.3~GHz. The detected masers 9.9 and 104.3~GHz and non-detected transitions 57.3 and 150.9~GHz belong to the series E~$J_{-1} \to (J-1)_{-2}$. At low and intermediate gas densities, our calculations predict population inversion for transitions at 28.3, 76.5 and 124.6~GHz belonging to the series E~$J_0 \to (J-1)_1$ (as yet non-detected), see Fig.~\ref{fig5}. Two detected transitions of E type methanol $4_2 \to 3_1$ 218.4~GHz and $5_2 \to 4_1$ 266.8~GHz have low optical depth in our calculations, $\vert \tau_{\mu=\text{1,\,l.c.}} \vert \lesssim 0.1$. Our results confirm the conclusions by \citet{Chen2019} -- these masers need a high gas density for pumping, $n_\mathrm{H_2} \sim 10^7$~cm$^{-3}$. In appendix~A we show that the calculated optical depth in these transitions is sensitive to the extrapolation of collisional rate coefficients to high temperatures. The list of methanol transitions is provided in Table~\ref{table:methtrans}.

The optical depth as a function of shock speed is presented in Fig.~\ref{fig6} for transitions of the series E~$J_2 \to J_1$ with even $J$. At $n_\mathrm{H,tot} = 2 \times 10^4$~cm$^{-3}$, the absolute value of the optical depth in all transitions of the E~$J_2 \to J_1$ series is low, and maser amplification is expected to be modest even at high aspect ratio $1/\mu = 10$. The most favourable gas density for the pumping of these masers is $n_\mathrm{H,tot} \approx 2 \times 10^5$~cm$^{-3}$. In this case, the optical depth is maximal for transitions with $J = 3,4,5$ and equals $\vert \tau_{\mu=\text{1,\,l.c.}} \vert \approx 1$. The optical depth in transitions of the E~$J_2 \to J_1$ series is lower than the optical depth in widespread maser transitions. 

The methanol concentration is higher at higher pre-shock gas density since we fix the methanol abundance relative to hydrogen nuclei at the start of shock wave simulations. However, the length of the shock region is inversely proportional to the pre-shock gas density, other things being equal. The cumulative effect is the lower absolute value of the optical depth at pre-shock gas density  $n_\text{H,tot} = 2 \times 10^5$~cm$^{-3}$ for transitions that are already effectively pumped at low gas density, e.g. for widespread masers 36.2 and 44.1~GHz, see Fig.~\ref{fig5}. The transitions that are relatively weak at low gas density become stronger at pre-shock gas density $n_\text{H,tot} = 2 \times 10^5$~cm$^{-3}$ -- E type methanol transitions near 25--30~GHz (E~$J_2 \to J_1$ series), 9.9, 104.3~GHz, and the A type methanol transition 23.4~GHz. These transitions belong to the group of 'rare' masers. It was shown earlier that these masers become bright in models with high gas density \citep{Sobolev2005,Leurini2018}.

At pre-shock gas density $n_\mathrm{H,tot} = 2 \times 10^6$~cm$^{-3}$, the methanol destruction at the shock front becomes important even at intermediate shock speeds. At shock speed 17.5~km~s$^{-1}$, the sputtering of icy mantles of dust grains is not complete -- the fraction of methanol sputtered to the gas phase is about 25 per cent. At this shock speed, the gas temperature at the shock front is already $T_\text{g} \approx 2000$~K, and methanol destruction in collisional dissociation reactions is effective. At shock speed $u_\mathrm{s} = 20$~km~s$^{-1}$, the gas-phase methanol is almost entirely destroyed in the hot gas at the shock front. As a result, the optical depth for all methanol masers $\vert \tau_{\mu=\text{1,\,l.c.}} \vert \lesssim 0.5$ at $n_\text{H,tot} = 2\times 10^6$~cm$^{-3}$, see Figs.~\ref{fig5} and \ref{fig6}. The range of allowable shock speeds for methanol maser operation is very narrow at such a high gas density.

\begin{table}
{\centering
\caption{Methanol maser transitions. \label{table:methtrans}}
\begin{tabular}{llll}
\hline
Transition & Frequency & Transition & Frequency \\
\hline
\multicolumn{2}{c}{A$^{+}$ $J_0 \to (J-1)_1$} & \multicolumn{2}{c}{E $J_{-1} \to (J-1)_0$} \\ [2pt]
$7_{0} \to 6_{1}$ & 44.1$^{*}$ &$4_{-1} \to 3_0$ & 36.2$^{*}$ \\[2pt]
$8_0 \to 7_1$ & 95.2$^{*}$     &$5_{-1} \to 4_0$ & 84.5$^{*}$ \\[2pt]
$9_0 \to 8_1$ & 146.6          &$6_{-1} \to 5_0$ & 132.9$^{*}$ \\[2pt]
$10_0 \to 9_1$ & 198.4         &$8_{-1} \to 7_0$ & 229.8$^{*}$ \\[2pt]
\multicolumn{2}{c}{A$^{+}$ $J_1 \to (J-1)_2$} & $9_{-1} \to 8_0$ & 278.3$^{*}$ \\[2pt]
$11_1 \to 10_2$ & 20.2         &\multicolumn{2}{c}{E $J_{-1} \to (J-1)_{-2}$} \\[2pt]
\multicolumn{2}{c}{A$^{-}$ $J_1 \to (J-1)_2$} &$9_{-1} \to 8_{-2}$ & 9.9$^{*}$ \\[2pt]
$10_1 \to 9_2$ & 23.4$^{*}$    &$10_{-1} \to 9_{-2}$ & 57.3 \\ [2pt]
$11_1 \to 10_2$ & 76.2         &$11_{-1} \to 10_{-2}$ & 104.3$^{*}$ \\[2pt]
\multicolumn{2}{c}{E $J_2 \to (J-1)_1$} &$12_{-1} \to 11_{-2}$ & 150.9 \\[2pt]
$4_2 \to 3_1$ & 218.4$^{*}$    & \multicolumn{2}{c}{E $J_0 \to (J-1)_1$} \\[2pt]
$5_2 \to 4_1$ & 266.8$^{*}$    & $4_0 \to 3_1$ & 28.3 \\[2pt]
 &                             & $5_0 \to 4_1$ & 76.5 \\[2pt]
 &                             & $6_0 \to 5_1$ & 124.6 \\[2pt]
\hline
\end{tabular} \par
}
\medskip
The methanol transitions listed have been detected (are marked by an asterisk) or have optical depth $\vert \tau_{\mu=\text{1,\,l.c.}} \vert > 0.5$ in at least one of the shock models. The transitions of the series E~$J_2 \to J_1$ are not presented here. Transition frequency is provided in GHz (the frequencies of methanol transitions are rounded, not truncated).
\end{table}

\subsection{Maser luminosities}
We find that the optical depth $\vert \tau_\text{sat} \vert$ at which widespread methanol masers become saturated is about 12--15, and the corresponding brightness temperature of maser emission $T_\mathrm{b,sat} \sim 10^6$--$10^7$~K. For the OH maser 1720~MHz, the parameter $\vert \tau_\text{sat} \vert$ is about 20 and $T_\mathrm{b,sat} \sim 10^9$--$10^{10}$~K. We take in our estimates the cosmic microwave background radiation as the background continuum radiation and a beaming angle of maser radiation $\Delta \Omega = 0.01$~sr. If we suggest that the shock wave is seen edge-on and the aspect ratio is $1/\mu \approx 10$, the optical depth in maser transitions is $\vert \tau_{\mu} \vert > 10 \vert \tau_{\mu=1} \vert$, and masers with $\vert \tau_{\mu=\text{1,\,l.c.}} \vert \gtrsim$ 1 are at least partly saturated. 

There are two tests of the degree of saturation of observed maser radiation: polarization of maser emission and deviation of the observed line profile of a maser from a Gaussian profile. The analysis of the OH maser spots near SNRs W28 and W44 indicated that OH masers are at least partially saturated \citep{Hoffman2005,Hoffman2005b}. Another test is based on a comparison between the observed brightness temperature and the maximal brightness temperature of the unsaturated maser inferred using the collisional pump model. \citet{Pihlstrom2014} reported Very Large Array observations of methanol masers in SNR W28. Here, we assume that the size of the bright spots in CH$_3$OH maser radiation is the same as that for OH masers -- the characteristic size of bright compact cores of OH maser emission near SNRs is $H \approx 10^{15}$~cm \citep{Hoffman2005,Hoffman2005b}. Based on the data by \citet{Pihlstrom2014}, the brightness temperatures of the most intense methanol masers in W28 are evaluated to be $T_\text{b} \approx 10^6$~K and $T_\text{b} \approx 10^5$~K for 44.1 and 36.2~GHz transitions, respectively. The brightness temperature of the 44.1~GHz maser is close to the saturation limit deduced from simulations.

\begin{figure*}
\includegraphics[width=150mm]{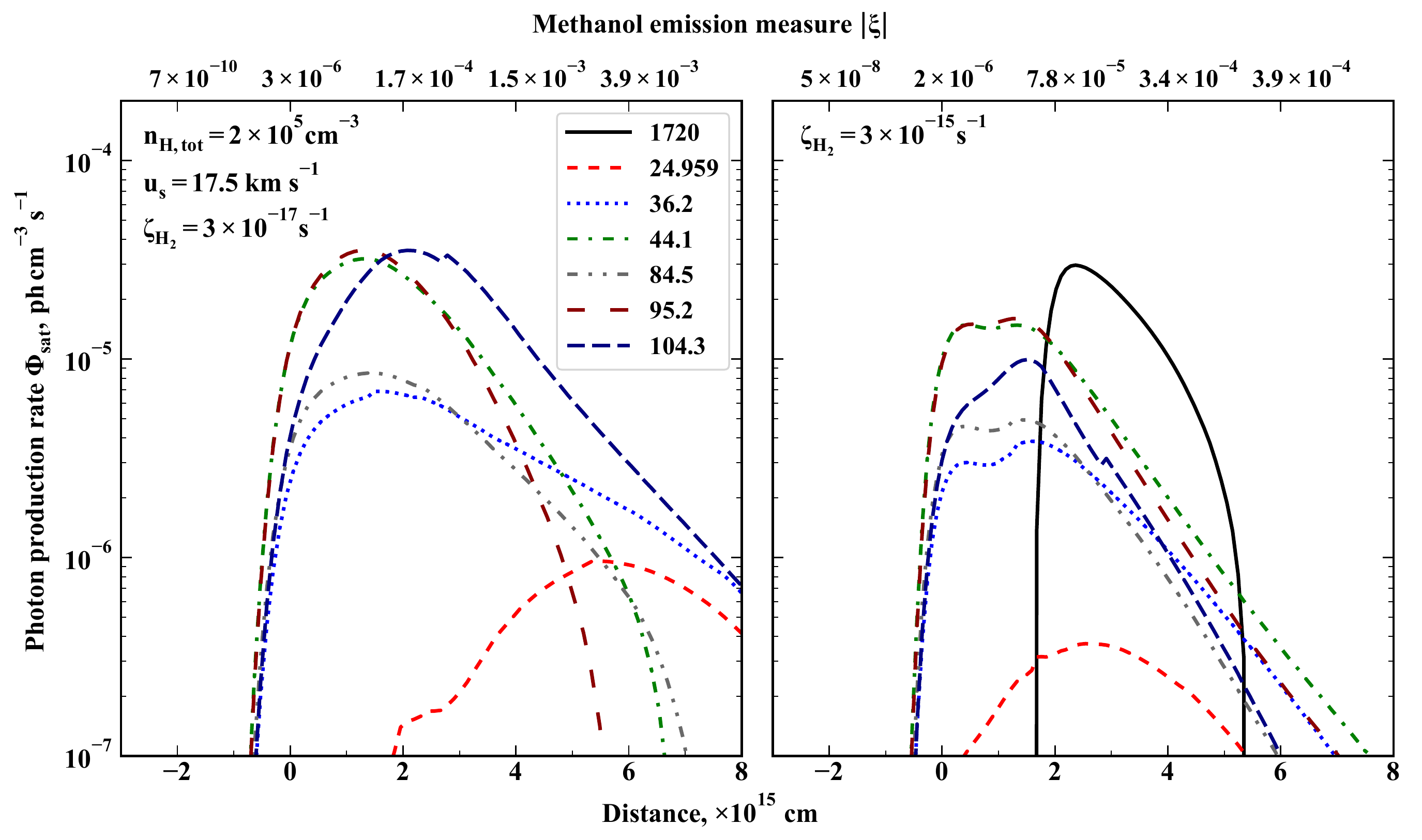}
\caption{Photon production rate for hydroxyl and methanol masers as a function of the distance in the shock. The bottom x-axis is for the distance along the gas flow, the zero is the point of the maximum temperature of the gas. The methanol emission measure $\vert \xi \vert$ is shown on the top x-axis. The model parameters are $n_\text{H,tot} = 2 \times 10^5$~cm$^{-3}$ and $u_\text{s} = 17.5$~km~s$^{-1}$. The results are presented for two values of cosmic ray ionization rate, $\zeta_\mathrm{H_2} = 3 \times 10^{-17}$~s$^{-1}$ (left) and $3 \times 10^{-15}$~s$^{-1}$ (right).}
\label{fig7}
\end{figure*}

Fig.~\ref{fig7} shows the photon production rate in saturation limit $\Phi_\text{sat}$ as a function of depth in the shock. The case of $n_\text{H,tot} = 2 \times 10^5$~cm$^{-3}$ and $u_\text{s} = 17.5$~km~s$^{-1}$ is considered. At cosmic ray ionization rate $\zeta_\mathrm{H_2} = 3 \times 10^{-17}$~s$^{-1}$, the hydroxyl maser 1720~MHz is absent. However, at $\zeta_\mathrm{H_2} = 3 \times 10^{-15}$~s$^{-1}$, the hydroxyl maser has the highest photon production rate, about $3 \times 10^{-5}$~ph~cm$^{-3}$~s$^{-1}$ at maximum. The photon production rate for 36.2 and 84.5~GHz transitions is 3--5 times lower than for 44.1 and 95.2~GHz transitions. The full width at half maximum of the photon production rate is about $2 \times 10^{15}$~cm for the OH maser and $3 \times 10^{15}$~cm for the CH$_3$OH masers. This length scale can be considered as the model prediction of the size of a fully saturated maser. The photon production rate in the lines of the methanol series E~$J_2 \to J_1$ is significantly lower than for widespread maser transitions -- analogous results were found by \citet{Leurini2016}. The photon production rate is proportional to the gas density. At shock model parameters $n_\text{H,tot} = 2 \times 10^4$~cm$^{-3}$, $\zeta_\mathrm{H_2} = 10^{-15}$~s$^{-1}$ and $u_\text{s} = 20$~km~s$^{-1}$, the photon production rate at maximum is about $\Phi_\text{sat} \approx 5 \times 10^{-7}$~ph~cm$^{-3}$~s$^{-1}$ for both hydroxyl and widespread methanol masers. 

The luminosity of a given maser transition can be crudely estimated as $L_\text{sat} \sim \Phi_\text{sat} H^3$, where $H$ is the size of the maser region. The strong maser emission is suggested to emanate predominantly in one direction within the solid angle $\Delta \Omega$. The isotropic photon luminosity of a maser in the saturated regime is 

\begin{equation}
\begin{array}{l}
\displaystyle
L_\text{is,th} = 4\pi L_\text{sat}/\Delta \Omega \approx \\
\displaystyle
\approx 10^{43} \left( \frac{\Phi_\text{sat}}{10^{-5} \,\text{ph}\, \text{cm}^{-3}\,\text{s}^{-1}} \right) \left( \frac{H}{10^{15}\,\text{cm}}\right)^3 \left(\frac{0.01\,\text{sr}}{\Delta\Omega} \right) \, \text{ph\,s}^{-1}.
\end{array}
\label{eq_lum_th}
\end{equation} 

\noindent
The isotropic photon luminosity of a maser feature can be estimated based on observational data:

\begin{equation}
L_\text{is,obs} \approx \frac{4\pi D^2 J \Delta \nu }{h\nu},
\end{equation} 

\noindent
where $J$ is the peak flux density of the maser feature in Jy, $\Delta \nu$ is the line width, $D$ is the distance to the source. \citet{Hoffman2005} presented Very Long Baseline Array (VLBA) observations of the 1720~MHz OH maser emission in SNR W28. For the component I of the most intense OH maser feature F39A in SNR W28 we obtain an estimate $L_\text{is,obs} \approx 5 \times 10^{42}$~ph~s$^{-1}$. Based on the data by \citet{Pihlstrom2014}, we estimate the isotropic photon luminosities of the most intense CH$_3$OH maser features in W28: $L_\text{is,obs} \approx 4 \times 10^{42}$~ph~s$^{-1}$ for the 44.1~GHz transition and $L_\text{is,obs} \approx 2 \times 10^{41}$~ph~s$^{-1}$ for the 36.2~GHz transition. The photon production rates and maser luminosities obtained in our calculations can reproduce the observations, see equation~(\ref{eq_lum_th}).

\begin{table}
\centering
\caption{Physical parameters in methanol maser sources according to simulations. \label{table:phys_param}}
\begin{tabular}{lll}
\hline
Parameter & $n_\text{H,tot} = $ & $n_\text{H,tot} = $ \\
 & $2 \times 10^4$~cm$^{-3}$& $2 \times 10^5$~cm$^{-3}$ \\
\hline
$n_\mathrm{H_2} \; \left[\text{cm}^{-3}\right]$ & $(2-7)\times 10^4$ & $(1.5-6)\times 10^5$\\ [5pt]
$H \; [\text{au}]$ & 1000 & 200 \\ [5pt]
$T_\text{g} \; [\text{K}]$ & $40-600$ & $100-1600$ \\ [5pt]
$n_\mathrm{CH_3OH} \; \left[\text{cm}^{-3}\right]$ & $0.5-1$ & $1-4$ \\ [5 pt]
$\vert\text{d}u/\text{d}z\vert \; \left[\text{cm~s}^{-1}\text{cm}^{-1}\right]$ & $10^{-11}-10^{-10}$ & $5 \times 10^{-11}- 5 \times 10^{-10}$\\ [5pt]
$N_\text{S} \; \left[\text{cm}^{-2}\right]$ & \multicolumn{2}{c}{$2 \times 10^{14}- 3 \times 10^{15}$} \\ [5pt]
$\vert \xi \vert$ & $2 \times 10^{-6} - 10^{-4}$ & $5 \times 10^{-6} - 10^{-3}$ \\
\hline
\end{tabular}
%\medskip 
\end{table}

\section{Discussion}
\subsection{Comparison with previous studies on Class I methanol masers}
The input parameters usually used in maser pumping simulations are gas density $n_\mathrm{H_2}$, maser region size $H$, gas temperature $T_\text{g}$, molecule concentration $n_\text{m}$, and velocity gradient $\text{d}u/\text{d}z$. Other parameters that are computed from these are column density of masing molecules $N_\text{S}$ along the Sobolev length $\Delta z_\text{S}$, and maser emission measure $\xi$. We provide the ranges of these parameters in the shock wave in Table~\ref{table:phys_param}. The results are given for the shock model with $u_\text{s} = 17.5$~km~s$^{-1}$, $\zeta_\mathrm{H_2} = 3 \times 10^{-17}$~s$^{-1}$ and for two pre-shock gas densities $n_\text{H,tot} = 2\times 10^4$ and $2 \times 10^5$~cm$^{-3}$. The ranges of parameters provided in Table~\ref{table:phys_param} correspond to the shock region where the gain in the 44.1 GHz transition is twice as low as the maximum value. Some of the physical parameters vary by more than an order of magnitude across the maser zone. 

Theoretical studies of the excitation of Class I methanol masers were done recently by \citet{McEwen2014,Nesterenok2016,Leurini2016}. These studies used a slab model for a warm gas--dust cloud where the physical parameters were assumed to be independent of the coordinates. The available data on collisional rate coefficients for the methanol molecule limit the gas temperature in such calculations to below 200~K. Here we show, that methanol is ejected to the gas phase at the shock peak where the gas temperature is high, $T_\text{g} \gtrsim 1000$~K. The contribution of the hot shocked gas to the total methanol emission may be non-negligible. In appendix \ref{app_ch3oh_excit} we discuss the possible error related to the absence of high temperature collisional data. 

According to our calculations, the absolute value of the optical depth in the 36.2~GHz transition is higher than that in the 44.1~GHz transition at low pre-shock gas density $n_\text{H,tot} = 2 \times 10^4$~cm$^{-3}$ and high cosmic ray ionization rate $\zeta_\mathrm{H_2} \gtrsim 10^{-15}$~s$^{-1}$, see Fig.~\ref{fig4}. In this case, the methanol column density along the Sobolev length is low, $N_\text{S} \lesssim 3 \times 10^{14}$~cm$^{-2}$. These conditions are more conducive to the pumping of the 36.2~GHz transition than the 44.1~GHz transition, see also fig.~1 in \citet{McEwen2014}. \citet{McEwen2014} also showed that the 36.2 GHz maser transition has a higher gain than the 44.1~GHz transition at high gas density, $n_\text{H,tot} \gtrsim 10^6$~cm$^{-3}$. Indeed, it is seen from Fig.~\ref{fig7} (left) that the photon production rate (and the gain) is higher for the 36.2~GHz transition in the shock tail. The significant changes in physical parameters across the shock region lead to the coexistence of different pumping regimes in one maser source, see also the discussion by \citet{Voronkov2006}. \citet{Leurini2016} suggested that bright Class I methanol masers are mainly high-density structures with high methanol maser emission measures close to the limits set by collisional quenching, $\vert \xi \vert \sim$ 0.1--1. These values of the parameter $\vert \xi \vert$ correspond to high gas density $n_\mathrm{H_2} \sim 10^7$--$10^8$~cm$^{-3}$. However, we have found that values of gas density in the maser zone in C-type shocks are lower than suggested by \citet{Leurini2016}. The methanol emission measure $\vert \xi \vert$ lies in the range $10^{-5}$--$10^{-3}$ in the maser zone at pre-shock gas density $n_\text{H,tot} = 2 \times 10^5$~cm$^{-3}$, see Fig.~\ref{fig7}. The absolute value of the emission measure $\vert \xi \vert$ increases to the shock end as the gas density increases and velocity gradient decreases with distance. 

\citet{Nesterenok2016} considered a simple one-dimensional model of a methanol maser, and used two methods for the calculation of the energy level populations of the molecule -- the accelerated lambda iteration method (ALI) and the LVG approximation. It was shown that the LVG approximation gives accurate results provided that the length scale of the change in physical parameters $l$ is larger than 5--10 lengths of the resonance region $\Delta z_\text{S}$. \citet{Nesterenok2016} showed that there is a significant discrepancy in the results of the ALI method and LVG approximation at $l < 3 \Delta z_\text{S}$. In the cooling gas flow behind the C-type shock front $l \approx (2-5) \Delta z_\text{S}$, and the LVG method may give inaccurate results. Radiation transfer methods such as ALI method take into account spatial variations of physical parameters. Accurate radiative transfer calculations on top of shock model simulations could be the next step in maser pumping modelling. We suppose that the results for maser transitions having the low optical depth will be more sensitive to radiative transfer model used.

It is worth noting that \citet{Hartquist1995} considered the coexistence of OH and CH$_3$OH Class II masers around ultracompact \ion{H}{II} regions. They suggested that methanol abundance is enhanced in the gas-phase by the evaporation of icy mantles of dust grains in weak shocks with speeds of few km~s$^{-1}$ propagating in dense gas, $n_\text{H,tot} = 10^7$~cm$^{-3}$. In our model, the dust temperature at the shock front is not high enough to guarantee the evaporation of icy mantles of dust grains even at the highest gas density in question: $T_\text{d} \lesssim 30$~K at $n_\text{H,tot} = 2 \times 10^6$~cm$^{-3}$ and $u_\text{s} \leq 15$~km~s$^{-1}$. The main mechanism of the release of adsorbed species in our shock model is sputtering. \citet{Yusef-Zadeh2013} studied widespread methanol emission at 36.2~GHz in the Central Molecular Zone (CMZ) -- molecular clouds in the inner few hundred pc of the Galaxy. They suggested a mechanism of methanol liberation from icy mantles of dust grains due to photo-desorption by cosmic ray-induced UV radiation. This mechanism has two disadvantages -- the methanol ice photo-desorption is inefficient since it is dominated by the desorption of photo-fragments \citep{Bertin2016}, and methanol has short destruction time-scale in the gas phase at high cosmic ray ionization rates.

\subsection{Coincidence of methanol and hydroxyl masers in SNRs}
Methanol emission originates in an extended region starting from the shock front, where the sputtering of icy grain mantles takes place, and ending in the shock tail, where the gas temperature falls below 30~K. The gain of widespread methanol masers has a wide maximum with a peak in the shock region where the gas temperature is a few hundred kelvin. The population inversion for the OH line 1720~MHz emerges when the gas temperature drops to about 300~K, and the maser gain has a maximum at about 70~K. Thus, the hydroxyl maser zone is more compact and is displaced to the shock end, see Fig.~\ref{fig3} and Fig.~\ref{fig7}. If the shock region is seen along the gas flow direction, the velocity difference between OH and CH$_3$OH emission is $\Delta u \lesssim$ 10--15~km~s$^{-1}$. The optical depth in the OH transition at 1720~MHz along the gas flow direction is $\vert \tau_{\mu=\text{1,\,l.c.}} \vert \lesssim 1.5$, see Fig.~\ref{fig4}. The line of sight must be perpendicular to the gas flow direction for generation of intense maser radiation. In this case, the shock is seen edge-on and the velocities of methanol and hydroxyl masers in the source must coincide. 

\citet{Pihlstrom2014} reported the detection of methanol maser emission at 36.2 and 44.1~GHz in SNR W28. In this source, two methanol maser detections at 36.2~GHz and two at 44.1~GHz are located within a primary beam distance (which equals about 1~pc) from the nearest OH masers (the locations of 44 GHz masers do not coincide with those of 36 GHz masers). The methanol masers are detected within a narrow velocity range between 6.8 and 8.2~km~s$^{-1}$ with respect to the local standard of rest (LSR), which is similar to the OH maser velocity at 7~km~s$^{-1}$ \citep{Pihlstrom2014}. The close agreement between velocities of methanol and OH masers argues for a physical association. The coexistence of methanol and hydroxyl masers in the same region allows one to infer physical conditions. The existence of OH maser emission sets the limit on the cosmic ray ionization rate $\zeta_\mathrm{H_2} \gtrsim 10^{-15}$~s$^{-1}$, while the existence of methanol maser emission does the same for the shock speed $u_\text{s} \lesssim 22.5$~km~s$^{-1}$. 

It seems natural to suppose that physical conditions are not uniform in a molecular cloud interacting with a SNR. In some parts of the SNR, the conditions may be favourable for the generation of hydroxyl masers and not of methanol masers, e.g. high shock speed $u_\text{s} > 20$~km~s$^{-1}$, or both low gas density $n_\text{H,tot} \lesssim 2 \times 10^4$~cm$^{-3}$ and high cosmic ray ionization rate. In contrast, in other parts of the SNR the conditions may be favourable for the generation of methanol masers -- relatively high gas density $n_\text{H,tot} \approx 2\times 10^5$~cm$^{-3}$ and low ionization rate $\zeta_\mathrm{H_2} < 10^{-15}$~s$^{-1}$. \citet{Sjouwerman2010} reported the interferometric observations of the Sgr~A complex in the 36.2 GHz transition of methanol with the Expanded Very Large Array. \citet{Pihlstrom2011}, and more recently \citet{McEwen2016b}, reported the survey of 44.1~GHz methanol maser emission in the Sgr~A region. The methanol and hydroxyl emission is concentrated in the north-eastern part of the Sgr~A East SNR where the SNR interacts with the 50~km~s$^{-1}$ molecular cloud. The closest separation between any methanol maser (36.2~GHz) and OH maser is about 0.25~pc, and OH masers have a higher mean LSR velocity compared to that for methanol masers \citep{Sjouwerman2010,Pihlstrom2011}. Spatial separation and velocity difference suggest that methanol and hydroxyl masers trace different shocks (or different locations in one shock region). Analogous picture is found in other SNRs -- the majority of methanol masers are found offset from 1720~MHz OH masers \citep{Pihlstrom2014,McEwen2016}. 

The 1720~MHz OH masers near SNRs were measured to have sizes of about 200--1000~au, with the most intense emission being generated by multiple closely spaced compact cores with sizes of about 50~au \citep{Hoffman2005,Hoffman2005b}. In particular, the maser feature F39A in SNR W28 has minimal and maximal sizes of about 200 and 700~au, respectively. The peaks observed in the VLBA image of this feature are separated in velocity by much less than one line width. \citet{Hoffman2005} concluded that the multiple peaks in the VLBA images are emission inhomogeneities within a single volume of maser gas, not dynamically distinct regions. This means that the size of the whole maser feature must be considered as an estimate of the size of the shock region where OH maser emission is generated. The sizes of OH maser features in SNRs indicate that pre-shock gas density lies in the range $2\times 10^4$--$2\times 10^5$~cm$^{-3}$ \citep{Nesterenok2020}. For methanol masers in SNRs, the angular size of maser spots may be established by interferometric observations in the future. Such observations in SNR W28 will allow to test the hypothesis of co-location of methanol and hydroxyl masers in the same shock region.

\subsection{Methanol masers in the Central Molecular Zone}
\citet{Zeng2020} studied the molecular emission towards G+0.693-0.03 -- a quiescent molecular cloud located within the Sgr~B2 star-forming complex in the CMZ. A total of 11 transitions of Class I methanol masers were detected towards this source. It was found that the 36.2~GHz maser emission is more than an order of magnitude brighter than the emission in the 44.1~GHz line. The masers are not associated with a star-formation process as there are no signposts of star formation observed towards G+0.693-0.03. The numerous Class I methanol masers at 36.2~GHz were found in another Galactic Centre cloud G0.253+0.016 -- a quiescent giant molecular cloud that also does not display advanced signatures of star formation \citep{Mills2015}. \citet{Cotton2016} conducted a 36.2~GHz line survey of the inner region of the Galactic Centre and found more than two thousands 36.2~GHz methanol maser sources. The molecular gas in the Galactic Centre is subject to widespread shocks and increased flux of cosmic ray particles and ionizing radiation. Cloud--cloud collision shocks and turbulent shocks in cloud interiors have been suggested to be the source of rich chemistry and excitation of Class I methanol masers in the CMZ \citep[e.g.,][]{Mehringer1997,Salii2002,RequenaTorres2006}. We have shown that there is a regime of maser pumping at relatively low gas density ($n_\mathrm{H,tot} \sim 2 \times 10^4$~cm$^{-3}$) and high cosmic ray ionization rate ($\zeta_\mathrm{H_2} \gtrsim 10^{-15}$~s$^{-1}$) where 36.2~GHz masers are stronger than 44.1~GHz masers, see Fig.~\ref{fig4}.

\cite{Ellingsen2017} used the Australia Telescope Compact Array (ATCA) to observe the Class I methanol maser emission at 36.2 and 44.1~GHz towards the starburst galaxy NGC~253. They found that methanol emission is located at a distance of a few hundred pc from the galactic nucleus, and the 44.1~GHz emission is two orders of magnitude weaker than the 36.2~GHz emission. The Class I methanol masers in NGC~253 seem to originate in large-scale cloud--cloud collision shocks \citep{Ellingsen2017}. Extremely high gas ionization rate may be a reason for the absence of methanol masers in the inner nuclear zone of NGC~253, as was pointed out by \cite{Ellingsen2017}. According to our calculations, a very high cosmic ray ionization rate, $\zeta_\mathrm{H_2} \gtrsim 10^{-14}$~s$^{-1}$, lowers the efficiency of methanol formation on dust grains and prevents its conservation in the gas phase. The Class I methanol maser emission was detected towards a number of galaxies \citep{Humire2020}. The relative brightness of 36.2~GHz emission is a common property of extragalactic methanol masers and masers towards the Galactic Centre.

\subsection{Comparison with observations of Class I methanol masers in bipolar outflows in star-forming regions}
\citet{Voronkov2014} presented an interferometric survey at both 36.2 and 44.1~GHz transitions of 71 southern Class I methanol masers using the ATCA. The methanol maser emission is associated with protostellar outflows, \ion{H}{II} regions, dark clouds, shocks traced by the infrared emission. \citet{Voronkov2014} investigated groups of emission components detected in both transitions, and found that A~type methanol emission at 44.1~GHz is stronger than E~type methanol emission at 36.2~GHz in the majority of cases. The median of the flux-density ratio for these transitions is 3. According to our calculations, the absolute value of the optical depth in the 44.1~GHz transition is larger than the optical depth in the 36.2~GHz transition at low cosmic ray ionization rate and/or at high gas density, see Figs.~\ref{fig4} and \ref{fig5}. The relative brightness of the A~type methanol transition 44.1~GHz may also be attributed to the isomer abundance ratio E/A < 1. If the spin temperature of methanol isomers reflects the cold environment of molecule formation at $T_\text{d} \approx 10$~K, the E/A isomer abundance ratio will be about 0.7 \citep{Holdship2019}. This E/A isomer abundance ratio is preserved in the post-shock gas if the relative abundance of A and E symmetry species does not change in the sputtering of icy mantles of dust grains. \citet{Holdship2019} found E/A isomer abundance ratio in the range 0.1--0.6 in seven protostellar outflow sources. The flux density ratio of methanol transitions 36.2 and 44.1~GHz is on average unity in SNRs -- in some locations the 36.2~GHz E~type methanol maser is brighter, the 44.1~GHz A~type methanol maser in others \citep{McEwen2016}. For example, the 36.2~GHz methanol masers detected in the north-eastern region of the Sgr~A East SNR are significantly brighter than the 44.1~GHz methanol masers: the flux density ratio between two maser transitions in this region is $\gtrsim 5$ \citep{McEwen2016b}. This can be explained only by the difference in physical conditions required to produce masers, as E/A isomer abundance ratio cannot be $ > 1$. 

\citet{McCarthy2018} made the high resolution observations of a large sample of Class I methanol masers in the 95.2~GHz transition using the ATCA. They found the flux density ratio between transitions 95.2 and 44.1~GHz to be equal to about 0.35 on average. \citet{Yang2020} reported a simultaneous 44.1 and 95.2~GHz Class I methanol maser survey using the telescopes of the Korean very long baseline interferometry network operating in single-dish mode. They found that the peak flux density ratio of the 95.2 and 44.1~GHz transitions ranges from 0.1 to 2.8, and the average value is 0.5. The masers 44.1 and 95.2~GHz belong to the same transition series A$^{+}$~$J_0 \to (J-1)_1$ and, hence, their relative brightness is not affected by the E/A isomer abundance ratio. According to our simulations, the relative brightness of the 44.1 and 95.2~GHz transitions depends on the gas density. At low pre-shock gas density $n_\text{H,tot} = 2 \times 10^4$~cm$^{-3}$, the absolute value of the optical depth in the 44.1~GHz transition is higher than that in the 95.2~GHz transition, see Fig.~\ref{fig5}. The transitions in question have close values of the optical depth and the photon production rate in saturation limit at high pre-shock gas density, $n_\text{H,tot} \geq 2 \times 10^5$~cm$^{-3}$, see Figs.~\ref{fig5} and \ref{fig7}.

\subsection{Absence of OH maser emission in 1720~MHz transition in bipolar outflows}
\citet{Litovchenko2012} suggested that the sites of Class I methanol masers in bipolar outflows in star-forming regions may be suitable for the generation of OH maser emission at 1720~MHz. \citet{DeWitt2014} undertook a search for the OH maser emission at 1720~MHz towards 97 Herbig--Haro objects. No OH maser emission at 1720~MHz associated with bipolar outflows was detected. \citet{DeWitt2014} pointed out that the size of the shock region in Herbig--Haro objects is not large enough to provide OH column densities needed for maser emission. Indeed, according to our calculations, the absolute value of the optical depth in the OH transition at 1720~MHz is relatively low even at high ionization rates, $\vert \tau_{\mu=\text{1,\,l.c.}} \vert \approx 1$, and the aspect ratio must be $1/\mu \approx 10$ to generate intense maser emission. The geometry and the size of the bow shock may not allow such high amplification paths. \citet{DeWitt2014} also considered whether the low ionization rate could be a reason for the absence of OH maser emission in Herbig--Haro objects. However, too low a threshold for the ionization rate required to produce a sufficient column of OH molecules ($\zeta_\text{H} = 10^{-16}$~s$^{-1}$) was used in their estimates. \citet{Bayandina2015,Bayandina2020} observed ground state OH transitions towards 20 Extended Green Objects that trace shocked gas in high-mass star formation regions. No evidence of an association between OH maser emission and shocked outflow gas traced by Class I methanol masers was found. 

According to our calculations, the high ionization rate of the molecular gas, $\zeta_\mathrm{H_2} \gtrsim 10^{-15}$~s$^{-1}$, is the necessary condition for the existence of OH masers at 1720~MHz in C-type shock waves. The gas ionization rates in molecular-cloud cores aside from the sources of ionizing radiation are much lower, of the order of $10^{-17}$--$10^{-16}$~s$^{-1}$ \citep{Dalgarno2006}. We suggest that an insufficient level of the gas ionization rate in the sites of Class I methanol masers in star-forming regions (that often are found at some distance from a central continuum source) could be a main reason for the absence of OH maser emission in these sites.

\subsection{Effect of the magnetic field strength and dust properties}
An analysis of molecular Zeeman data suggests that the magnetic field strength can have a wide range of values at any given density of interstellar cloud \citep{Crutcher2010}. The maximal magnetic field strength is adopted in our simulations: parameter $\beta = 1$ in equation (\ref{eq_magn_field}). Moreover, for the sake of simplicity we assume that the magnetic field is parallel to the shock front. The length of the shock region is in direct ratio with the transverse magnetic field strength. The narrower the region where kinetic energy dissipates is, the higher the peak temperature of the gas at the shock front, which in turn affects molecular abundances. For the species that are produced in the cooling post-shock gas (OH molecule), a low magnetic field strength leads to a small absolute value of the optical depth in molecular transitions \citep{Nesterenok2020}.  

In the interstellar gas, dust grains have a distribution of sizes, and grains with a fixed size have a distribution of charge. Moreover, PAH molecules may be present in the pre-shock gas that increases the specific surface area of dust. The larger the specific surface area of dust is, the smaller the length of the shock region. The small charged grains remain coupled to the ion fluid, while large grains are partially decoupled by collisions with the neutrals. Grain--grain collisions lead to the shattering of grains into small fragments that leads to an increase in the specific surface area of dust grains -- this effect is not taken into account in our model. Grain shattering is found to be important in the dynamics of C-type shocks at $n_\text{H,tot} \gtrsim 10^5$~cm$^{-3}$ \citep{Guillet2011,Anderl2013}. As a result of grain shattering, the C-type shock becomes hotter and thinner. 

At high pre-shock gas density, $n_\text{H,tot} = 2 \times 10^6$~cm$^{-3}$, gas-phase methanol is effectively destroyed at the shock front at shock speeds just above the sputtering threshold. Grain shuttering will result in a higher gas temperature at the shock front, and the fraction of methanol surviving the passage of the shock wave will be lower. A magnetic field strength lower than that adopted in the calculations will result in an analogous effect. 
% Hoffman 2005a section 4.4, Balsara et al. 2001

\section{Conclusions}
The evolution of molecular abundances and collisional pumping of OH and CH$_3$OH masers in C-type shock waves have been considered. The calculations predict the population inversion for all the detected Class I methanol masers. The following conclusions can be drawn from the conducted analysis: 

(i) Methanol masers in the shock wave can exist for a narrow range of shock speeds, $15 \leq u_\text{s} \leq 22.5$~km~s$^{-1}$. At shock speeds $u_\text{s} \gtrsim 25$~km~s$^{-1}$, the collisional dissociation of gas-phase methanol takes place at the shock front. Pre-shock gas densities $n_\text{H,tot} = 2 \times 10^4$--$2 \times 10^5$~cm$^{-3}$ are favourable for the pumping of methanol maser transitions. At high pre-shock gas density $n_\text{H,tot} = 2 \times 10^6$~cm$^{-3}$ and at any shock speed above the threshold of grain mantle sputtering, the methanol abundance in the gas phase is low due to collisional dissociation.

(ii) The higher the cosmic ray ionization rate is, the higher the OH concentration in the post-shock region, and the higher the absolute value of the optical depth in the OH maser transition at 1720~MHz. At the same time, the methanol column density decreases with increasing cosmic ray ionization rate. At low cosmic ray ionization rate, $\zeta_\mathrm{H_2} \lesssim 10^{-16}$~s$^{-1}$, the optical depth in CH$_3$OH transitions is maximal. At the same time, the OH maser emission is absent -- the optical depth along the gas flow in the 1720~MHz transition is $\vert \tau_{\mu=1} \vert < 0.1$ for such a cosmic ray ionization rate. A low ionization rate may be a reason for the absence of OH maser emission at 1720~MHz from locations of Class I methanol masers in bipolar outflows. 

(iii) The OH and CH$_3$OH masers can operate simultaneously at high cosmic ray ionization rate $\zeta_\mathrm{H_2} \approx 10^{-15}$--$3 \times 10^{-15}$~s$^{-1}$, and shock speed $u_\text{s} \approx$ 15--20~km~s$^{-1}$. The optical depth along the gas flow is $\vert \tau_{\mu=\text{1,\,l.c.}} \vert \approx$ 0.5--1 in this case. The line of sight should be perpendicular to the gas flow direction (the shock is seen edge-on) in order to generate strong maser emission. Simultaneous observations of both OH and CH$_3$OH masers in one object may be used to set strict constraints on the physical parameters of such an object.

(iv) At pre-shock gas density $n_\text{H,tot} = 2 \times 10^4$~cm$^{-3}$ and high cosmic ray ionization rate $\zeta_\mathrm{H_2} \gtrsim 10^{-15}$~s$^{-1}$, the absolute value of the optical depth for the methanol maser transition at 36.2~GHz is higher than that for the 44.1~GHz transition. The relative brightness of 36.2~GHz masers over 44.1~GHz masers in SNRs and in molecular clouds of the Galactic Centre region may be attributed to the high level of gas ionization rate. The absolute value of the optical depth of the 44.1~GHz transition is larger than that of the 36.2~GHz transition at low ionization rate and/or high gas density.

(v) The post-shock region where the gas temperature descends below a few hundred kelvin makes the main contribution to methanol and hydroxyl emission. Calculations with collisional rate coefficients of methanol extrapolated to high gas temperatures have shown that the effect of the extrapolation on the optical depth is small for most of the methanol masers. However, we have found that the optical depth in weak methanol masers, e.g. E type transitions of 218.4 and 266.8~GHz, is sensitive to extrapolation of collisional rate coefficients. 

(vi) The physical parameters vary by more than an order of magnitude across the post-shock region where methanol maser emission is generated. Different pump regimes may coexist in the maser zone. Some uncertainty of the calculated maser parameters may be caused by the LVG approximation applied for radiative transfer calculations.

\section*{Acknowledgements}
This work was supported by RSF grant 21-72-20020. I acknowledge the anonymous referee for helpful comments that have significantly improved this paper.

\section*{Data availability}
The data underlying this article will be shared on reasonable request to the corresponding author.

\bibliographystyle{mnras}
\bibliography{../../interstellar_medium_references,../../chemistry_references}

\begin{thebibliography}{}
\makeatletter
\relax
\def\mn@urlcharsother{\let\do\@makeother \do\$\do\&\do\#\do\^\do\_\do\%\do\~}
\def\mn@doi{\begingroup\mn@urlcharsother \@ifnextchar [ {\mn@doi@}
  {\mn@doi@[]}}
\def\mn@doi@[#1]#2{\def\@tempa{#1}\ifx\@tempa\@empty \href
  {http://dx.doi.org/#2} {doi:#2}\else \href {http://dx.doi.org/#2} {#1}\fi
  \endgroup}
\def\mn@eprint#1#2{\mn@eprint@#1:#2::\@nil}
\def\mn@eprint@arXiv#1{\href {http://arxiv.org/abs/#1} {{\tt arXiv:#1}}}
\def\mn@eprint@dblp#1{\href {http://dblp.uni-trier.de/rec/bibtex/#1.xml}
  {dblp:#1}}
\def\mn@eprint@#1:#2:#3:#4\@nil{\def\@tempa {#1}\def\@tempb {#2}\def\@tempc
  {#3}\ifx \@tempc \@empty \let \@tempc \@tempb \let \@tempb \@tempa \fi \ifx
  \@tempb \@empty \def\@tempb {arXiv}\fi \@ifundefined
  {mn@eprint@\@tempb}{\@tempb:\@tempc}{\expandafter \expandafter \csname
  mn@eprint@\@tempb\endcsname \expandafter{\@tempc}}}

\bibitem[\protect\citeauthoryear{{Anderl}, {Guillet}, {Pineau des For{\^e}ts}
  \& {Flower}}{{Anderl} et~al.}{2013}]{Anderl2013}
{Anderl} S.,  {Guillet} V.,  {Pineau des For{\^e}ts} G.,   {Flower} D.~R.,
  2013, \mn@doi [\aap] {10.1051/0004-6361/201321399}, \href
  {http://adsabs.harvard.edu/abs/2013A%26A...556A..69A} {556, A69}

\bibitem[\protect\citeauthoryear{{Ascenzi}, {Cernuto}, {Balucani}, {Tosi},
  {Ceccarelli}, {Martini}  \& {Pirani}}{{Ascenzi} et~al.}{2019}]{Ascenzi2019}
{Ascenzi} D.,  {Cernuto} A.,  {Balucani} N.,  {Tosi} P.,  {Ceccarelli} C.,
  {Martini} L.~M.,   {Pirani} F.,  2019, \mn@doi [\aap]
  {10.1051/0004-6361/201834585}, \href
  {https://ui.adsabs.harvard.edu/abs/2019A&A...625A..72A} {625, A72}

\bibitem[\protect\citeauthoryear{{Batrla}, {Matthews}, {Menten}  \&
  {Walmsley}}{{Batrla} et~al.}{1987}]{Batrla1987}
{Batrla} W.,  {Matthews} H.~E.,  {Menten} K.~M.,   {Walmsley} C.~M.,  1987,
  \mn@doi [\nat] {10.1038/326049a0}, \href
  {http://adsabs.harvard.edu/abs/1987Natur.326...49B} {326, 49}

\bibitem[\protect\citeauthoryear{{Baulch} et~al.,}{{Baulch}
  et~al.}{2005}]{Baulch2005}
{Baulch} D.~L.,  et~al., 2005, \mn@doi
  [\JournalOfPhysicalChemicalReferenceData] {10.1063/1.1748524}, \href
  {http://adsabs.harvard.edu/abs/2005JPCRD..34..757B} {34, 757}

\bibitem[\protect\citeauthoryear{{Bayandina}, {Val'tts}  \&
  {Kurtz}}{{Bayandina} et~al.}{2015}]{Bayandina2015}
{Bayandina} O.~S.,  {Val'tts} I.~E.,   {Kurtz} S.~E.,  2015, \mn@doi
  [\AstronomyReports] {10.1134/S1063772915110025}, \href
  {https://ui.adsabs.harvard.edu/abs/2015ARep...59..998B} {59, 998}

\bibitem[\protect\citeauthoryear{{Bayandina}, {Colom}, {Kurtz}, {Rudnitskij},
  {Shakhvorostova}  \& {Val'tts}}{{Bayandina} et~al.}{2020}]{Bayandina2020}
{Bayandina} O.~S.,  {Colom} P.,  {Kurtz} S.~E.,  {Rudnitskij} G.~M.,
  {Shakhvorostova} N.~N.,   {Val'tts} I.~E.,  2020, \mn@doi [\mnras]
  {10.1093/mnras/staa2885}, \href
  {https://ui.adsabs.harvard.edu/abs/2020MNRAS.499.3961B} {499, 3961}

\bibitem[\protect\citeauthoryear{{Bertin} et~al.,}{{Bertin}
  et~al.}{2016}]{Bertin2016}
{Bertin} M.,  et~al., 2016, \mn@doi [\apjl] {10.3847/2041-8205/817/2/L12},
  \href {https://ui.adsabs.harvard.edu/abs/2016ApJ...817L..12B} {817, L12}

\bibitem[\protect\citeauthoryear{{Beuther} et~al.,}{{Beuther}
  et~al.}{2019}]{Beuther2019}
{Beuther} H.,  et~al., 2019, \mn@doi [\aap] {10.1051/0004-6361/201935936},
  \href {https://ui.adsabs.harvard.edu/abs/2019A&A...628A..90B} {628, A90}

\bibitem[\protect\citeauthoryear{{Boogert}, {Gerakines}  \&
  {Whittet}}{{Boogert} et~al.}{2015}]{Boogert2015}
{Boogert} A.~C.~A.,  {Gerakines} P.~A.,   {Whittet} D.~C.~B.,  2015, \mn@doi
  [\araa] {10.1146/annurev-astro-082214-122348}, \href
  {http://adsabs.harvard.edu/abs/2015ARA%26A..53..541B} {53, 541}

\bibitem[\protect\citeauthoryear{{Bossion}, {Scribano}, {Lique}  \&
  {Parlant}}{{Bossion} et~al.}{2018}]{Bossion2018}
{Bossion} D.,  {Scribano} Y.,  {Lique} F.,   {Parlant} G.,  2018, \mn@doi
  [\mnras] {10.1093/mnras/sty2089}, \href
  {http://adsabs.harvard.edu/abs/2018MNRAS.480.3718B} {480, 3718}

\bibitem[\protect\citeauthoryear{{Breen}, {Contreras}, {Dawson}, {Ellingsen},
  {Voronkov}  \& {McCarthy}}{{Breen} et~al.}{2019}]{Breen2019}
{Breen} S.~L.,  {Contreras} Y.,  {Dawson} J.~R.,  {Ellingsen} S.~P.,
  {Voronkov} M.~A.,   {McCarthy} T.~P.,  2019, \mn@doi [\mnras]
  {10.1093/mnras/stz192}, \href
  {https://ui.adsabs.harvard.edu/abs/2019MNRAS.484.5072B} {484, 5072}

\bibitem[\protect\citeauthoryear{{Brogan} et~al.,}{{Brogan}
  et~al.}{2013}]{Brogan2013}
{Brogan} C.~L.,  et~al., 2013, \mn@doi [\apj] {10.1088/0004-637X/771/2/91},
  \href {https://ui.adsabs.harvard.edu/abs/2013ApJ...771...91B} {771, 91}

\bibitem[\protect\citeauthoryear{{Burdyuzha} \& {Varshalovich}}{{Burdyuzha} \&
  {Varshalovich}}{1973}]{Burdyuzha1973}
{Burdyuzha} V.~V.,  {Varshalovich} D.~A.,  1973, \sovast, \href
  {https://ui.adsabs.harvard.edu/abs/1973SvA....17..308B} {17, 308}

\bibitem[\protect\citeauthoryear{{Ceballos}, {Garcia}  \&
  {Lagan{\`a}}}{{Ceballos} et~al.}{2002}]{Ceballos2002}
{Ceballos} A.,  {Garcia} E.,   {Lagan{\`a}} A.,  2002, \mn@doi
  [\JournalOfPhysicalChemicalReferenceData] {10.1063/1.1461830}, \href
  {http://adsabs.harvard.edu/abs/2002JPCRD..31..371C} {31, 371}

\bibitem[\protect\citeauthoryear{{Chabot}, {B{\'e}roff}, {Gratier}, {Jallat}
  \& {Wakelam}}{{Chabot} et~al.}{2013}]{Chabot2013}
{Chabot} M.,  {B{\'e}roff} K.,  {Gratier} P.,  {Jallat} A.,   {Wakelam} V.,
  2013, \mn@doi [\apj] {10.1088/0004-637X/771/2/90}, \href
  {http://adsabs.harvard.edu/abs/2013ApJ...771...90C} {771, 90}

\bibitem[\protect\citeauthoryear{{Chen}, {Ellingsen}, {Ren}, {Sobolev},
  {Parfenov}  \& {Shen}}{{Chen} et~al.}{2019}]{Chen2019}
{Chen} X.,  {Ellingsen} S.~P.,  {Ren} Z.-Y.,  {Sobolev} A.~M.,  {Parfenov} S.,
   {Shen} Z.-Q.,  2019, \mn@doi [\apj] {10.3847/1538-4357/ab1078}, 877, 90

\bibitem[\protect\citeauthoryear{{Cotton} \& {Yusef-Zadeh}}{{Cotton} \&
  {Yusef-Zadeh}}{2016}]{Cotton2016}
{Cotton} W.~D.,  {Yusef-Zadeh} F.,  2016, \mn@doi [\apjs]
  {10.3847/0067-0049/227/1/10}, \href
  {https://ui.adsabs.harvard.edu/abs/2016ApJS..227...10C} {227, 10}

\bibitem[\protect\citeauthoryear{{Cragg}, {Johns}, {Godfrey}  \&
  {Brown}}{{Cragg} et~al.}{1992}]{Cragg1992}
{Cragg} D.~M.,  {Johns} K.~P.,  {Godfrey} P.~D.,   {Brown} R.~D.,  1992,
  \mnras, \href {http://adsabs.harvard.edu/abs/1992MNRAS.259..203C} {259, 203}

\bibitem[\protect\citeauthoryear{{Cragg}, {Sobolev}  \& {Godfrey}}{{Cragg}
  et~al.}{2002}]{Cragg2002}
{Cragg} D.~M.,  {Sobolev} A.~M.,   {Godfrey} P.~D.,  2002, \mn@doi [\mnras]
  {10.1046/j.1365-8711.2002.05226.x}, \href
  {http://adsabs.harvard.edu/abs/2002MNRAS.331..521C} {331, 521}

\bibitem[\protect\citeauthoryear{{Crutcher}}{{Crutcher}}{1999}]{Crutcher1999}
{Crutcher} R.~M.,  1999, \mn@doi [\apj] {10.1086/307483}, \href
  {http://adsabs.harvard.edu/abs/1999ApJ...520..706C} {520, 706}

\bibitem[\protect\citeauthoryear{{Crutcher}, {Wandelt}, {Heiles}, {Falgarone}
  \& {Troland}}{{Crutcher} et~al.}{2010}]{Crutcher2010}
{Crutcher} R.~M.,  {Wandelt} B.,  {Heiles} C.,  {Falgarone} E.,   {Troland}
  T.~H.,  2010, \mn@doi [\apj] {10.1088/0004-637X/725/1/466}, \href
  {https://ui.adsabs.harvard.edu/abs/2010ApJ...725..466C} {725, 466}

\bibitem[\protect\citeauthoryear{{Dalgarno}}{{Dalgarno}}{2006}]{Dalgarno2006}
{Dalgarno} A.,  2006, \mn@doi [\ProceedingsOfTheNationalAcademyOfScience]
  {10.1073/pnas.0602117103}, \href
  {http://adsabs.harvard.edu/abs/2006PNAS..10312269D} {103, 12269}

\bibitem[\protect\citeauthoryear{{Doel}, {Gray}  \& {Field}}{{Doel}
  et~al.}{1990}]{Doel1990}
{Doel} R.~C.,  {Gray} M.~D.,   {Field} D.,  1990, \mnras, \href
  {https://ui.adsabs.harvard.edu/abs/1990MNRAS.244..504D} {244, 504}

\bibitem[\protect\citeauthoryear{{Dudorov}}{{Dudorov}}{1991}]{Dudorov1991}
{Dudorov} A.~E.,  1991, \sovast, \href
  {https://ui.adsabs.harvard.edu/abs/1991SvA....35..342D} {35, 342}

\bibitem[\protect\citeauthoryear{{Elitzur}, {Hollenbach}  \& {McKee}}{{Elitzur}
  et~al.}{1989}]{Elitzur1989}
{Elitzur} M.,  {Hollenbach} D.~J.,   {McKee} C.~F.,  1989, \mn@doi [\apj]
  {10.1086/168080}, \href
  {https://ui.adsabs.harvard.edu/abs/1989ApJ...346..983E} {346, 983}

\bibitem[\protect\citeauthoryear{{Ellingsen}, {Chen}, {Breen}  \&
  {Qiao}}{{Ellingsen} et~al.}{2017}]{Ellingsen2017}
{Ellingsen} S.~P.,  {Chen} X.,  {Breen} S.~L.,   {Qiao} H.~H.,  2017, \mn@doi
  [\mnras] {10.1093/mnras/stx2076}, \href
  {https://ui.adsabs.harvard.edu/abs/2017MNRAS.472..604E} {472, 604}

\bibitem[\protect\citeauthoryear{{Faure}, {Vuitton}, {Thissen}, {Wiesenfeld}
  \& {Dutuit}}{{Faure} et~al.}{2010}]{Faure2010}
{Faure} A.,  {Vuitton} V.,  {Thissen} R.,  {Wiesenfeld} L.,   {Dutuit} O.,
  2010, \mn@doi [Faraday Discussions] {10.1039/c003908j}, \href
  {http://adsabs.harvard.edu/abs/2010FaDi..147..337F} {147, 337}

\bibitem[\protect\citeauthoryear{{Flower} \& {Gusdorf}}{{Flower} \&
  {Gusdorf}}{2009}]{Flower2009}
{Flower} D.~R.,  {Gusdorf} A.,  2009, \mn@doi [\mnras]
  {10.1111/j.1365-2966.2009.14569.x}, \href
  {http://adsabs.harvard.edu/abs/2009MNRAS.395..234F} {395, 234}

\bibitem[\protect\citeauthoryear{{Flower} \& {Pineau des For{\^e}ts}}{{Flower}
  \& {Pineau des For{\^e}ts}}{2003}]{Flower2003}
{Flower} D.~R.,  {Pineau des For{\^e}ts} G.,  2003, \mn@doi [\mnras]
  {10.1046/j.1365-8711.2003.06716.x}, \href
  {http://adsabs.harvard.edu/abs/2003MNRAS.343..390F} {343, 390}

\bibitem[\protect\citeauthoryear{{Flower}, {Pineau Des For{\^e}ts}  \&
  {Walmsley}}{{Flower} et~al.}{2006}]{Flower2006}
{Flower} D.~R.,  {Pineau Des For{\^e}ts} G.,   {Walmsley} C.~M.,  2006, \mn@doi
  [\aap] {10.1051/0004-6361:20054246}, \href
  {http://adsabs.harvard.edu/abs/2006A%26A...449..621F} {449, 621}

\bibitem[\protect\citeauthoryear{{Flower}, {Pineau Des For{\^e}ts}  \&
  {Rabli}}{{Flower} et~al.}{2010}]{Flower2010b}
{Flower} D.~R.,  {Pineau Des For{\^e}ts} G.,   {Rabli} D.,  2010, \mn@doi
  [\mnras] {10.1111/j.1365-2966.2010.17501.x}, \href
  {http://adsabs.harvard.edu/abs/2010MNRAS.409...29F} {409, 29}

\bibitem[\protect\citeauthoryear{{Frail} \& {Mitchell}}{{Frail} \&
  {Mitchell}}{1998}]{Frail1998}
{Frail} D.~A.,  {Mitchell} G.~F.,  1998, \mn@doi [\apj] {10.1086/306452}, \href
  {https://ui.adsabs.harvard.edu/abs/1998ApJ...508..690F} {508, 690}

\bibitem[\protect\citeauthoryear{{Frail}, {Goss}, {Reynoso}, {Giacani}, {Green}
   \& {Otrupcek}}{{Frail} et~al.}{1996}]{Frail1996}
{Frail} D.~A.,  {Goss} W.~M.,  {Reynoso} E.~M.,  {Giacani} E.~B.,  {Green}
  A.~J.,   {Otrupcek} R.,  1996, \mn@doi [\aj] {10.1086/117904}, \href
  {https://ui.adsabs.harvard.edu/abs/1996AJ....111.1651F} {111, 1651}

\bibitem[\protect\citeauthoryear{{Gao}, {Zheng}, {Fern\'andez-Ramos}, {Truhlar}
   \& {Xu}}{{Gao} et~al.}{2018}]{Gao2018}
{Gao} L.~G.,  {Zheng} J.,  {Fern\'andez-Ramos} A.,  {Truhlar} D.~G.,   {Xu} X.,
   2018, \mn@doi [\JournalOfTheAmericanChemicalSociety] {10.1021/jacs.7b12773},
  140, 2906

\bibitem[\protect\citeauthoryear{{Garrod}, {Park}, {Caselli}  \&
  {Herbst}}{{Garrod} et~al.}{2006}]{Garrod2006}
{Garrod} R.,  {Park} I.~H.,  {Caselli} P.,   {Herbst} E.,  2006, \mn@doi
  [Faraday Discussions] {10.1039/b516202e}, \href
  {http://adsabs.harvard.edu/abs/2006FaDi..133...51G} {133, 51}

\bibitem[\protect\citeauthoryear{{Garrod}, {Wakelam}  \& {Herbst}}{{Garrod}
  et~al.}{2007}]{Garrod2007}
{Garrod} R.~T.,  {Wakelam} V.,   {Herbst} E.,  2007, \mn@doi [\aap]
  {10.1051/0004-6361:20066704}, \href
  {http://adsabs.harvard.edu/abs/2007A%26A...467.1103G} {467, 1103}

\bibitem[\protect\citeauthoryear{{Gonz{\'a}lez-Lezana} \&
  {Honvault}}{{Gonz{\'a}lez-Lezana} \& {Honvault}}{2017}]{Gonzalez-Lezana2017}
{Gonz{\'a}lez-Lezana} T.,  {Honvault} P.,  2017, \mn@doi [\mnras]
  {10.1093/mnras/stx192}, \href
  {http://adsabs.harvard.edu/abs/2017MNRAS.467.1294G} {467, 1294}

\bibitem[\protect\citeauthoryear{{Gordon} et~al.,}{{Gordon}
  et~al.}{2017}]{Gordon2017}
{Gordon} I.~E.,  et~al., 2017, \mn@doi [\jqsrt]
  {https://doi.org/10.1016/j.jqsrt.2017.06.038}, 203, 3

\bibitem[\protect\citeauthoryear{{Goto}, {Vasyunin}, {Giuliano},
  {Jim{\'e}nez-Serra}, {Caselli}, {Rom{\'a}n-Z{\'u}{\~n}iga}  \&
  {Alves}}{{Goto} et~al.}{2021}]{Goto2021}
{Goto} M.,  {Vasyunin} A.~I.,  {Giuliano} B.~M.,  {Jim{\'e}nez-Serra} I.,
  {Caselli} P.,  {Rom{\'a}n-Z{\'u}{\~n}iga} C.~G.,   {Alves} J.,  2021, \mn@doi
  [\aap] {https://doi.org/10.1051/0004-6361/201936385}, \href
  {https://ui.adsabs.harvard.edu/abs/2020arXiv201210883G} {651, A53}

\bibitem[\protect\citeauthoryear{{Guillet}, {Pineau Des For{\^e}ts}  \&
  {Jones}}{{Guillet} et~al.}{2011}]{Guillet2011}
{Guillet} V.,  {Pineau Des For{\^e}ts} G.,   {Jones} A.~P.,  2011, \mn@doi
  [\aap] {10.1051/0004-6361/201015973}, \href
  {http://adsabs.harvard.edu/abs/2011A%26A...527A.123G} {527, A123}

\bibitem[\protect\citeauthoryear{{Gusdorf}, {Cabrit}, {Flower}  \& {Pineau Des
  For{\^e}ts}}{{Gusdorf} et~al.}{2008}]{Gusdorf2008}
{Gusdorf} A.,  {Cabrit} S.,  {Flower} D.~R.,   {Pineau Des For{\^e}ts} G.,
  2008, \mn@doi [\aap] {10.1051/0004-6361:20078900}, \href
  {http://adsabs.harvard.edu/abs/2008A%26A...482..809G} {482, 809}

\bibitem[\protect\citeauthoryear{{Hartquist}, {Menten}, {Lepp}  \&
  {Dalgarno}}{{Hartquist} et~al.}{1995}]{Hartquist1995}
{Hartquist} T.~W.,  {Menten} K.~M.,  {Lepp} S.,   {Dalgarno} A.,  1995, \mnras,
  \href {http://adsabs.harvard.edu/abs/1995MNRAS.272..184H} {272, 184}

\bibitem[\protect\citeauthoryear{{Heays}, {Bosman}  \& {van Dishoeck}}{{Heays}
  et~al.}{2017}]{Heays2017}
{Heays} A.~N.,  {Bosman} A.~D.,   {van Dishoeck} E.~F.,  2017, \mn@doi [\aap]
  {10.1051/0004-6361/201628742}, \href
  {http://adsabs.harvard.edu/abs/2017A%26A...602A.105H} {602, A105}

\bibitem[\protect\citeauthoryear{{Hewitt}, {Yusef-Zadeh}  \& {Wardle}}{{Hewitt}
  et~al.}{2008}]{Hewitt2008}
{Hewitt} J.~W.,  {Yusef-Zadeh} F.,   {Wardle} M.,  2008, \mn@doi [\apj]
  {10.1086/588652}, \href
  {https://ui.adsabs.harvard.edu/abs/2008ApJ...683..189H} {683, 189}

\bibitem[\protect\citeauthoryear{{Hoffman}, {Goss}, {Brogan}  \&
  {Claussen}}{{Hoffman} et~al.}{2005a}]{Hoffman2005}
{Hoffman} I.~M.,  {Goss} W.~M.,  {Brogan} C.~L.,   {Claussen} M.~J.,  2005a,
  \mn@doi [\apj] {10.1086/427018}, \href
  {https://ui.adsabs.harvard.edu/abs/2005ApJ...620..257H} {620, 257}

\bibitem[\protect\citeauthoryear{{Hoffman}, {Goss}, {Brogan}  \&
  {Claussen}}{{Hoffman} et~al.}{2005b}]{Hoffman2005b}
{Hoffman} I.~M.,  {Goss} W.~M.,  {Brogan} C.~L.,   {Claussen} M.~J.,  2005b,
  \mn@doi [\apj] {10.1086/430419}, \href
  {https://ui.adsabs.harvard.edu/abs/2005ApJ...627..803H} {627, 803}

\bibitem[\protect\citeauthoryear{{Holdship} et~al.,}{{Holdship}
  et~al.}{2019}]{Holdship2019}
{Holdship} J.,  et~al., 2019, \mn@doi [\apj] {10.3847/1538-4357/ab1f8f}, \href
  {https://ui.adsabs.harvard.edu/abs/2019ApJ...880..138H} {880, 138}

\bibitem[\protect\citeauthoryear{{Humire} et~al.,}{{Humire}
  et~al.}{2020}]{Humire2020}
{Humire} P.~K.,  et~al., 2020, \mn@doi [\aap] {10.1051/0004-6361/201936330},
  \href {https://ui.adsabs.harvard.edu/abs/2020A&A...633A.106H} {633, A106}

\bibitem[\protect\citeauthoryear{{Hummer} \& {Rybicki}}{{Hummer} \&
  {Rybicki}}{1985}]{Hummer1985}
{Hummer} D.~G.,  {Rybicki} G.~B.,  1985, \mn@doi [\apj] {10.1086/163232}, \href
  {http://adsabs.harvard.edu/abs/1985ApJ...293..258H} {293, 258}

\bibitem[\protect\citeauthoryear{{Kalv{\= a}ns}}{{Kalv{\=
  a}ns}}{2018}]{Kalvans2018}
{Kalv{\= a}ns} J.,  2018, \mn@doi [\mnras] {10.1093/mnras/sty1172}, \href
  {http://adsabs.harvard.edu/abs/2018MNRAS.478.2753K} {478, 2753}

\bibitem[\protect\citeauthoryear{{Kurtz}, {Hofner}  \& {{\'A}lvarez}}{{Kurtz}
  et~al.}{2004}]{Kurtz2004}
{Kurtz} S.,  {Hofner} P.,   {{\'A}lvarez} C.~V.,  2004, \mn@doi [\apjs]
  {10.1086/423956}, \href
  {https://ui.adsabs.harvard.edu/abs/2004ApJS..155..149K} {155, 149}

\bibitem[\protect\citeauthoryear{{Ladeyschikov}, {Bayandina}  \&
  {Sobolev}}{{Ladeyschikov} et~al.}{2019}]{Ladeyschikov2019}
{Ladeyschikov} D.~A.,  {Bayandina} O.~S.,   {Sobolev} A.~M.,  2019, \mn@doi
  [\aj] {10.3847/1538-3881/ab4b4c}, \href
  {https://ui.adsabs.harvard.edu/abs/2019AJ....158..233L} {158, 233}

\bibitem[\protect\citeauthoryear{{Leurini} \& {Menten}}{{Leurini} \&
  {Menten}}{2018}]{Leurini2018}
{Leurini} S.,  {Menten} K.~M.,  2018, in {Tarchi} A.,  {Reid} M.~J.,
  {Castangia} P.,  eds,  IAU Symposium Vol. 336, Astrophysical Masers:
  Unlocking the Mysteries of the Universe. Cambridge University Press, pp
  17--22, \mn@doi{10.1017/S1743921317010705}

\bibitem[\protect\citeauthoryear{{Leurini}, {Menten}  \& {Walmsley}}{{Leurini}
  et~al.}{2016}]{Leurini2016}
{Leurini} S.,  {Menten} K.~M.,   {Walmsley} C.~M.,  2016, \mn@doi [\aap]
  {10.1051/0004-6361/201527974}, \href
  {https://ui.adsabs.harvard.edu/abs/2016A&A...592A..31L} {592, A31}

\bibitem[\protect\citeauthoryear{{Li}, {Xu}, {Chen}, {Lu}, {Sun}, {Du}  \&
  {Shen}}{{Li} et~al.}{2017}]{Li2017}
{Li} Y.-J.,  {Xu} Y.,  {Chen} X.,  {Lu} D.-R.,  {Sun} Y.,  {Du} X.-Y.,   {Shen}
  Z.-Q.,  2017, \mn@doi [\ResearchInAstronomyAndAstrophysics]
  {10.1088/1674-4527/17/12/125}, \href
  {https://ui.adsabs.harvard.edu/abs/2017RAA....17..125L} {17, 125}

\bibitem[\protect\citeauthoryear{{Lique}}{{Lique}}{2015}]{Lique2015}
{Lique} F.,  2015, \mn@doi [\mnras] {10.1093/mnras/stv1683}, \href
  {http://adsabs.harvard.edu/abs/2015MNRAS.453..810L} {453, 810}

\bibitem[\protect\citeauthoryear{{Lique}, {Honvault}  \& {Faure}}{{Lique}
  et~al.}{2014}]{Lique2014}
{Lique} F.,  {Honvault} P.,   {Faure} A.,  2014, \mn@doi
  [\InternationalReviewsInPhysicalChemistry] {10.1080/0144235X.2014.897443},
  33, 125

\bibitem[\protect\citeauthoryear{{Litovchenko}, {Alakoz}, {Val'tts}  \&
  {Larionov}}{{Litovchenko} et~al.}{2011}]{Litovchenko2011}
{Litovchenko} I.~D.,  {Alakoz} A.~V.,  {Val'tts} I.~E.,   {Larionov} G.~M.,
  2011, \mn@doi [\AstronomyReports] {10.1134/S1063772911110059}, \href
  {https://ui.adsabs.harvard.edu/abs/2011ARep...55..978L} {55, 978}

\bibitem[\protect\citeauthoryear{{Litovchenko} et~al.,}{{Litovchenko}
  et~al.}{2012}]{Litovchenko2012}
{Litovchenko} I.~D.,  et~al., 2012, \mn@doi [\AstronomyReports]
  {10.1134/S1063772912060042}, \href
  {https://ui.adsabs.harvard.edu/abs/2012ARep...56..536L} {56, 536}

\bibitem[\protect\citeauthoryear{{Litvak}}{{Litvak}}{1969}]{Litvak1969}
{Litvak} M.~M.,  1969, \mn@doi [\apj] {10.1086/149982}, \href
  {https://ui.adsabs.harvard.edu/abs/1969ApJ...156..471L} {156, 471}

\bibitem[\protect\citeauthoryear{{Liu}, {Chen}  \& {Du}}{{Liu}
  et~al.}{2020}]{Liu2020}
{Liu} C.,  {Chen} X.,   {Du} F.,  2020, \mn@doi [\apj]
  {10.3847/1538-4357/aba758}, \href
  {https://ui.adsabs.harvard.edu/abs/2020ApJ...899...92L} {899, 92}

\bibitem[\protect\citeauthoryear{{Marinakis}, {Kalugina}  \&
  {Lique}}{{Marinakis} et~al.}{2016}]{Marinakis2016}
{Marinakis} S.,  {Kalugina} Y.,   {Lique} F.,  2016, \mn@doi
  [\EuropeanPhysicalJournalD] {10.1140/epjd/e2016-70068-x}, 70, 97

\bibitem[\protect\citeauthoryear{{Marinakis}, {Kalugina}, {K{\l}os}  \&
  {Lique}}{{Marinakis} et~al.}{2019}]{Marinakis2019}
{Marinakis} S.,  {Kalugina} Y.,  {K{\l}os} J.,   {Lique} F.,  2019, \mn@doi
  [\aap] {10.1051/0004-6361/201936170}, \href
  {https://ui.adsabs.harvard.edu/abs/2019A&A...629A.130M} {629, A130}

\bibitem[\protect\citeauthoryear{{Martin}, {Keogh}  \& {Mandy}}{{Martin}
  et~al.}{1998}]{Martin1998}
{Martin} P.~G.,  {Keogh} W.~J.,   {Mandy} M.~E.,  1998, \mn@doi [\apj]
  {10.1086/305665}, \href {http://adsabs.harvard.edu/abs/1998ApJ...499..793M}
  {499, 793}

\bibitem[\protect\citeauthoryear{{McCarthy}, {Ellingsen}, {Voronkov}  \&
  {Cim{\`o}}}{{McCarthy} et~al.}{2018}]{McCarthy2018}
{McCarthy} T.~P.,  {Ellingsen} S.~P.,  {Voronkov} M.~A.,   {Cim{\`o}} G.,
  2018, \mn@doi [\mnras] {10.1093/mnras/sty694}, \href
  {https://ui.adsabs.harvard.edu/abs/2018MNRAS.477..507M} {477, 507}

\bibitem[\protect\citeauthoryear{{McElroy}, {Walsh}, {Markwick}, {Cordiner},
  {Smith}  \& {Millar}}{{McElroy} et~al.}{2013}]{McElroy2013}
{McElroy} D.,  {Walsh} C.,  {Markwick} A.~J.,  {Cordiner} M.~A.,  {Smith} K.,
  {Millar} T.~J.,  2013, \mn@doi [\aap] {10.1051/0004-6361/201220465}, \href
  {http://adsabs.harvard.edu/abs/2013A%26A...550A..36M} {550, A36}

\bibitem[\protect\citeauthoryear{{McEwen}, {Pihlstr{\"o}m}  \&
  {Sjouwerman}}{{McEwen} et~al.}{2014}]{McEwen2014}
{McEwen} B.~C.,  {Pihlstr{\"o}m} Y.~M.,   {Sjouwerman} L.~O.,  2014, \mn@doi
  [\apj] {10.1088/0004-637X/793/2/133}, \href
  {http://adsabs.harvard.edu/abs/2014ApJ...793..133M} {793, 133}

\bibitem[\protect\citeauthoryear{{McEwen}, {Pihlstr{\"o}m}  \&
  {Sjouwerman}}{{McEwen} et~al.}{2016a}]{McEwen2016}
{McEwen} B.~C.,  {Pihlstr{\"o}m} Y.~M.,   {Sjouwerman} L.~O.,  2016a, \mn@doi
  [\apj] {10.3847/0004-637X/826/2/189}, \href
  {http://adsabs.harvard.edu/abs/2016ApJ...826..189M} {826, 189}

\bibitem[\protect\citeauthoryear{{McEwen}, {Sjouwerman}  \&
  {Pihlstr{\"o}m}}{{McEwen} et~al.}{2016b}]{McEwen2016b}
{McEwen} B.~C.,  {Sjouwerman} L.~O.,   {Pihlstr{\"o}m} Y.~M.,  2016b, \mn@doi
  [\apj] {10.3847/0004-637X/832/2/129}, \href
  {https://ui.adsabs.harvard.edu/abs/2016ApJ...832..129M} {832, 129}

\bibitem[\protect\citeauthoryear{{Mehringer} \& {Menten}}{{Mehringer} \&
  {Menten}}{1997}]{Mehringer1997}
{Mehringer} D.~M.,  {Menten} K.~M.,  1997, \mn@doi [\apj] {10.1086/303454},
  \href {https://ui.adsabs.harvard.edu/abs/1997ApJ...474..346M} {474, 346}

\bibitem[\protect\citeauthoryear{{Mekhtiev}, {Godfrey}  \& {Hougen}}{{Mekhtiev}
  et~al.}{1999}]{Mekhtiev1999}
{Mekhtiev} M.~A.,  {Godfrey} P.~D.,   {Hougen} J.~T.,  1999, \mn@doi
  [\JournalOfMolecularSpectroscopy] {10.1006/jmsp.1998.7782}, \href
  {http://adsabs.harvard.edu/abs/1999JMoSp.194..171M} {194, 171}

\bibitem[\protect\citeauthoryear{{Menten}}{{Menten}}{1991}]{Menten1991b}
{Menten} K.,  1991, in {Haschick} A.~D.,  {Ho} P.~T.~P.,  eds,  Astronomical
  Society of the Pacific Conference Series Vol. 16, Atoms, Ions and Molecules:
  New Results in Spectral Line Astrophysics. p.~119

\bibitem[\protect\citeauthoryear{{Mills}, {Butterfield}, {Ludovici}, {Lang},
  {Ott}, {Morris}  \& {Schmitz}}{{Mills} et~al.}{2015}]{Mills2015}
{Mills} E.~A.~C.,  {Butterfield} N.,  {Ludovici} D.~A.,  {Lang} C.~C.,  {Ott}
  J.,  {Morris} M.~R.,   {Schmitz} S.,  2015, \mn@doi [\apj]
  {10.1088/0004-637X/805/1/72}, \href
  {https://ui.adsabs.harvard.edu/abs/2015ApJ...805...72M} {805, 72}

\bibitem[\protect\citeauthoryear{{Minissale}, {Congiu}, {Manic{\`o}},
  {Pirronello}  \& {Dulieu}}{{Minissale} et~al.}{2013}]{Minissale2013}
{Minissale} M.,  {Congiu} E.,  {Manic{\`o}} G.,  {Pirronello} V.,   {Dulieu}
  F.,  2013, \mn@doi [\aap] {10.1051/0004-6361/201321453}, \href
  {https://ui.adsabs.harvard.edu/abs/2013A&A...559A..49M} {559, A49}

\bibitem[\protect\citeauthoryear{{Nesterenok}}{{Nesterenok}}{2016}]{Nesterenok2016}
{Nesterenok} A.~V.,  2016, \mn@doi [\mnras] {10.1093/mnras/stv2594}, \href
  {http://adsabs.harvard.edu/abs/2016MNRAS.455.3978N} {455, 3978}

\bibitem[\protect\citeauthoryear{{Nesterenok}}{{Nesterenok}}{2018}]{Nesterenok2018}
{Nesterenok} A.~V.,  2018, \mn@doi [\apss] {10.1007/s10509-018-3370-6}, \href
  {http://adsabs.harvard.edu/abs/2018Ap%26SS.363..151N} {363, 151}

\bibitem[\protect\citeauthoryear{{Nesterenok}}{{Nesterenok}}{2020a}]{Nesterenok2020}
{Nesterenok} A.~V.,  2020a, \mn@doi [\AstronomyLetters]
  {10.1134/S1063773720070075}, 46, 449

\bibitem[\protect\citeauthoryear{{Nesterenok}}{{Nesterenok}}{2020b}]{Nesterenok2020b}
{Nesterenok} A.~V.,  2020b, \mn@doi [\JournalOfPhysicsConferenceSeries]
  {10.1088/1742-6596/1697/1/012001}, 1697, 012001

\bibitem[\protect\citeauthoryear{{Nesterenok}, {Bossion}, {Scribano}  \&
  {Lique}}{{Nesterenok} et~al.}{2019}]{Nesterenok2019}
{Nesterenok} A.~V.,  {Bossion} D.,  {Scribano} Y.,   {Lique} F.,  2019, \mn@doi
  [\mnras] {10.1093/mnras/stz2441}, \href
  {https://ui.adsabs.harvard.edu/abs/2019MNRAS.489.4520N} {489, 4520}

\bibitem[\protect\citeauthoryear{{Neufeld} \& {Melnick}}{{Neufeld} \&
  {Melnick}}{1991}]{Neufeld1991}
{Neufeld} D.~A.,  {Melnick} G.~J.,  1991, \mn@doi [\apj] {10.1086/169685},
  \href {http://adsabs.harvard.edu/abs/1991ApJ...368..215N} {368, 215}

\bibitem[\protect\citeauthoryear{{Ng}}{{Ng}}{1974}]{Ng1974}
{Ng} K.-C.,  1974, \mn@doi [\jcp] {10.1063/1.1682399}, \href
  {http://adsabs.harvard.edu/abs/1974JChPh..61.2680N} {61, 2680}

\bibitem[\protect\citeauthoryear{{Offer}, {van Hemert}  \& {van
  Dishoeck}}{{Offer} et~al.}{1994}]{Offer1994}
{Offer} A.,  {van Hemert} M.,   {van Dishoeck} E.,  1994, \mn@doi
  [\JournalOfChemicalPhysics] {10.1063/1.466950}, 100, 362

\bibitem[\protect\citeauthoryear{{Padovani}, {Galli}, {Ivlev}, {Caselli}  \&
  {Ferrara}}{{Padovani} et~al.}{2018}]{Padovani2018}
{Padovani} M.,  {Galli} D.,  {Ivlev} A.~V.,  {Caselli} P.,   {Ferrara} A.,
  2018, \mn@doi [\aap] {10.1051/0004-6361/201834008}, \href
  {http://adsabs.harvard.edu/abs/2018A%26A...619A.144P} {619, A144}

\bibitem[\protect\citeauthoryear{{Pagani} et~al.,}{{Pagani}
  et~al.}{2009}]{Pagani2009}
{Pagani} L.,  et~al., 2009, \mn@doi [\aap] {10.1051/0004-6361:200810587}, \href
  {http://adsabs.harvard.edu/abs/2009A%26A...494..623P} {494, 623}

\bibitem[\protect\citeauthoryear{{Pagani}, {Lesaffre}, {Jorfi}, {Honvault},
  {Gonz{\'a}lez-Lezana}  \& {Faure}}{{Pagani} et~al.}{2013}]{Pagani2013}
{Pagani} L.,  {Lesaffre} P.,  {Jorfi} M.,  {Honvault} P.,
  {Gonz{\'a}lez-Lezana} T.,   {Faure} A.,  2013, \mn@doi [\aap]
  {10.1051/0004-6361/201117161}, \href
  {http://adsabs.harvard.edu/abs/2013A%26A...551A..38P} {551, A38}

\bibitem[\protect\citeauthoryear{{Palau} et~al.,}{{Palau}
  et~al.}{2017}]{Palau2017}
{Palau} A.,  et~al., 2017, \mn@doi [\mnras] {10.1093/mnras/stx004}, \href
  {http://adsabs.harvard.edu/abs/2017MNRAS.467.2723P} {467, 2723}

\bibitem[\protect\citeauthoryear{{Pastchenko} \& {Slysh}}{{Pastchenko} \&
  {Slysh}}{1974}]{Pastchenko1974}
{Pastchenko} M.~I.,  {Slysh} V.~I.,  1974, \aap, \href
  {https://ui.adsabs.harvard.edu/abs/1974A&A....35..153P} {35, 153}

\bibitem[\protect\citeauthoryear{{Pavlakis} \& {Kylafis}}{{Pavlakis} \&
  {Kylafis}}{1996}]{Pavlakis1996b}
{Pavlakis} K.~G.,  {Kylafis} N.~D.,  1996, \mn@doi [\apj] {10.1086/177605},
  \href {https://ui.adsabs.harvard.edu/abs/1996ApJ...467..300P} {467, 300}

\bibitem[\protect\citeauthoryear{{Penteado}, {Walsh}  \& {Cuppen}}{{Penteado}
  et~al.}{2017}]{Penteado2017}
{Penteado} E.~M.,  {Walsh} C.,   {Cuppen} H.~M.,  2017, \mn@doi [\apj]
  {10.3847/1538-4357/aa78f9}, \href
  {http://adsabs.harvard.edu/abs/2017ApJ...844...71P} {844, 71}

\bibitem[\protect\citeauthoryear{{Pihlstr{\"o}m}, {Sjouwerman}  \&
  {Fish}}{{Pihlstr{\"o}m} et~al.}{2011}]{Pihlstrom2011}
{Pihlstr{\"o}m} Y.~M.,  {Sjouwerman} L.~O.,   {Fish} V.~L.,  2011, \mn@doi
  [\apjl] {10.1088/2041-8205/739/1/L21}, \href
  {https://ui.adsabs.harvard.edu/abs/2011ApJ...739L..21P} {739, L21}

\bibitem[\protect\citeauthoryear{{Pihlstr{\"o}m}, {Sjouwerman}, {Frail},
  {Claussen}, {Mesler}  \& {McEwen}}{{Pihlstr{\"o}m}
  et~al.}{2014}]{Pihlstrom2014}
{Pihlstr{\"o}m} Y.~M.,  {Sjouwerman} L.~O.,  {Frail} D.~A.,  {Claussen} M.~J.,
  {Mesler} R.~A.,   {McEwen} B.~C.,  2014, \mn@doi [\aj]
  {10.1088/0004-6256/147/4/73}, \href
  {https://ui.adsabs.harvard.edu/abs/2014AJ....147...73P} {147, 73}

\bibitem[\protect\citeauthoryear{{Qiao} et~al.,}{{Qiao}
  et~al.}{2020}]{Qiao2020}
{Qiao} H.-H.,  et~al., 2020, \mn@doi [\apjs] {10.3847/1538-4365/ab655d}, \href
  {https://ui.adsabs.harvard.edu/abs/2020ApJS..247....5Q} {247, 5}

\bibitem[\protect\citeauthoryear{{Rabli} \& {Flower}}{{Rabli} \&
  {Flower}}{2010a}]{Rabli2010}
{Rabli} D.,  {Flower} D.~R.,  2010a, \mn@doi [\mnras]
  {10.1111/j.1365-2966.2010.16240.x}, \href
  {http://adsabs.harvard.edu/abs/2010MNRAS.403.2033R} {403, 2033}

\bibitem[\protect\citeauthoryear{{Rabli} \& {Flower}}{{Rabli} \&
  {Flower}}{2010b}]{Rabli2010b}
{Rabli} D.,  {Flower} D.~R.,  2010b, \mn@doi [\mnras]
  {10.1111/j.1365-2966.2010.16671.x}, \href
  {http://adsabs.harvard.edu/abs/2010MNRAS.406...95R} {406, 95}

\bibitem[\protect\citeauthoryear{{Rabli} \& {Flower}}{{Rabli} \&
  {Flower}}{2011}]{Rabli2011}
{Rabli} D.,  {Flower} D.~R.,  2011, \mn@doi [\mnras]
  {10.1111/j.1365-2966.2010.17842.x}, \href
  {http://adsabs.harvard.edu/abs/2011MNRAS.411.2093R} {411, 2093}

\bibitem[\protect\citeauthoryear{{Requena-Torres}, {Mart{\'{\i}}n-Pintado},
  {Rodr{\'{\i}}guez-Franco}, {Mart{\'{\i}}n}, {Rodr{\'{\i}}guez-Fern{\'a}ndez}
  \& {de Vicente}}{{Requena-Torres} et~al.}{2006}]{RequenaTorres2006}
{Requena-Torres} M.~A.,  {Mart{\'{\i}}n-Pintado} J.,  {Rodr{\'{\i}}guez-Franco}
  A.,  {Mart{\'{\i}}n} S.,  {Rodr{\'{\i}}guez-Fern{\'a}ndez} N.~J.,   {de
  Vicente} P.,  2006, \mn@doi [\aap] {10.1051/0004-6361:20065190}, \href
  {http://adsabs.harvard.edu/abs/2006A%26A...455..971R} {455, 971}

\bibitem[\protect\citeauthoryear{{Roberts}, {Rawlings}, {Viti}  \&
  {Williams}}{{Roberts} et~al.}{2007}]{Roberts2007}
{Roberts} J.~F.,  {Rawlings} J.~M.~C.,  {Viti} S.,   {Williams} D.~A.,  2007,
  \mn@doi [\mnras] {10.1111/j.1365-2966.2007.12402.x}, \href
  {http://adsabs.harvard.edu/abs/2007MNRAS.382..733R} {382, 733}

\bibitem[\protect\citeauthoryear{{Ruaud}, {Wakelam}  \& {Hersant}}{{Ruaud}
  et~al.}{2016}]{Ruaud2016}
{Ruaud} M.,  {Wakelam} V.,   {Hersant} F.,  2016, \mn@doi [\mnras]
  {10.1093/mnras/stw887}, \href
  {http://adsabs.harvard.edu/abs/2016MNRAS.459.3756R} {459, 3756}

\bibitem[\protect\citeauthoryear{{Ruffle} \& {Herbst}}{{Ruffle} \&
  {Herbst}}{2001}]{Ruffle2001}
{Ruffle} D.~P.,  {Herbst} E.,  2001, \mn@doi [\mnras]
  {10.1046/j.1365-8711.2001.04178.x}, \href
  {http://adsabs.harvard.edu/abs/2001MNRAS.322..770R} {322, 770}

\bibitem[\protect\citeauthoryear{{Salii}, {Sobolev}  \& {Kalinina}}{{Salii}
  et~al.}{2002}]{Salii2002}
{Salii} S.~V.,  {Sobolev} A.~M.,   {Kalinina} N.~D.,  2002, \mn@doi
  [\AstronomyReports] {10.1134/1.1529254}, \href
  {https://ui.adsabs.harvard.edu/abs/2002ARep...46..955S} {46, 955}

\bibitem[\protect\citeauthoryear{{Sch{\"o}ier}, {van der Tak}, {van Dishoeck}
  \& {Black}}{{Sch{\"o}ier} et~al.}{2005}]{Schoier2005}
{Sch{\"o}ier} F.~L.,  {van der Tak} F.~F.~S.,  {van Dishoeck} E.~F.,   {Black}
  J.~H.,  2005, \mn@doi [\aap] {10.1051/0004-6361:20041729}, \href
  {https://ui.adsabs.harvard.edu/abs/2005A&A...432..369S} {432, 369}

\bibitem[\protect\citeauthoryear{{Shaw}, {Ferland}  \& {Ploeckinger}}{{Shaw}
  et~al.}{2020}]{Shaw2020}
{Shaw} G.,  {Ferland} G.~J.,   {Ploeckinger} S.,  2020, \mn@doi [Research Notes
  of the American Astronomical Society] {10.3847/2515-5172/ab97ae}, \href
  {https://ui.adsabs.harvard.edu/abs/2020RNAAS...4...78S} {4, 78}

\bibitem[\protect\citeauthoryear{{Shingledecker}, {Bergner}, {Le Gal},
  {{\"O}berg}, {Hincelin}  \& {Herbst}}{{Shingledecker}
  et~al.}{2016}]{Shingledecker2016}
{Shingledecker} C.~N.,  {Bergner} J.~B.,  {Le Gal} R.,  {{\"O}berg} K.~I.,
  {Hincelin} U.,   {Herbst} E.,  2016, \mn@doi [\apj]
  {10.3847/0004-637X/830/2/151}, \href
  {http://adsabs.harvard.edu/abs/2016ApJ...830..151S} {830, 151}

\bibitem[\protect\citeauthoryear{{Shingledecker}, {Tennis}, {Le Gal}  \&
  {Herbst}}{{Shingledecker} et~al.}{2018}]{Shingledecker2018}
{Shingledecker} C.~N.,  {Tennis} J.,  {Le Gal} R.,   {Herbst} E.,  2018,
  \mn@doi [\apj] {10.3847/1538-4357/aac5ee}, \href
  {https://ui.adsabs.harvard.edu/abs/2018ApJ...861...20S} {861, 20}

\bibitem[\protect\citeauthoryear{{Sjouwerman}, {Pihlstr{\"o}m}  \&
  {Fish}}{{Sjouwerman} et~al.}{2010}]{Sjouwerman2010}
{Sjouwerman} L.~O.,  {Pihlstr{\"o}m} Y.~M.,   {Fish} V.~L.,  2010, \mn@doi
  [\apjl] {10.1088/2041-8205/710/2/L111}, \href
  {https://ui.adsabs.harvard.edu/abs/2010ApJ...710L.111S} {710, L111}

\bibitem[\protect\citeauthoryear{{Sobolev}}{{Sobolev}}{1957}]{Sobolev1957}
{Sobolev} V.~V.,  1957, \sovast, \href
  {http://adsabs.harvard.edu/abs/1957SvA.....1..678S} {1, 678}

\bibitem[\protect\citeauthoryear{{Sobolev}}{{Sobolev}}{1992}]{Sobolev1992}
{Sobolev} A.~M.,  1992, \sovast, \href
  {https://ui.adsabs.harvard.edu/abs/1992SvA....36..590S} {36, 590}

\bibitem[\protect\citeauthoryear{{Sobolev}, {Ostrovskii}, {Kirsanova},
  {Shelemei}, {Voronkov}  \& {Malyshev}}{{Sobolev} et~al.}{2005}]{Sobolev2005}
{Sobolev} A.~M.,  {Ostrovskii} A.~B.,  {Kirsanova} M.~S.,  {Shelemei} O.~V.,
  {Voronkov} M.~A.,   {Malyshev} A.~V.,  2005, in {Cesaroni} R.,  {Felli} M.,
  {Churchwell} E.,   {Walmsley} M.,  eds,  IAU Symposium Vol. 227, Massive Star
  Birth: A Crossroads of Astrophysics. Cambridge University Press, pp 174--179,
  \mn@doi{10.1017/S1743921305004503}

\bibitem[\protect\citeauthoryear{{Suutarinen}, {Kristensen}, {Mottram},
  {Fraser}  \& {van Dishoeck}}{{Suutarinen} et~al.}{2014}]{Suutarinen2014}
{Suutarinen} A.~N.,  {Kristensen} L.~E.,  {Mottram} J.~C.,  {Fraser} H.~J.,
  {van Dishoeck} E.~F.,  2014, \mn@doi [\mnras] {10.1093/mnras/stu406}, \href
  {http://adsabs.harvard.edu/abs/2014MNRAS.440.1844S} {440, 1844}

\bibitem[\protect\citeauthoryear{{Szczepanski}, {Ho}, {Haschick}  \&
  {Baan}}{{Szczepanski} et~al.}{1989}]{Szczepanski1989}
{Szczepanski} J.~C.,  {Ho} P.~T.~P.,  {Haschick} A.~D.,   {Baan} W.~A.,  1989,
  in {Morris} M.,  ed.,  IAU Symposium Vol. 136, The Center of the Galaxy.
  Kluwer Academic Publishers, Dordrecht, p.~383

\bibitem[\protect\citeauthoryear{{Towner}, {Brogan}, {Hunter}, {Cyganowski},
  {McGuire}, {Indebetouw}, {Friesen}  \& {Chandler}}{{Towner}
  et~al.}{2017}]{Towner2017}
{Towner} A.~P.~M.,  {Brogan} C.~L.,  {Hunter} T.~R.,  {Cyganowski} C.~J.,
  {McGuire} B.~A.,  {Indebetouw} R.,  {Friesen} R.~K.,   {Chandler} C.~J.,
  2017, \mn@doi [\apjs] {10.3847/1538-4365/aa73d8}, \href
  {https://ui.adsabs.harvard.edu/abs/2017ApJS..230...22T} {230, 22}

\bibitem[\protect\citeauthoryear{{Trevisan} \& {Tennyson}}{{Trevisan} \&
  {Tennyson}}{2002}]{Trevisan2002}
{Trevisan} C.~S.,  {Tennyson} J.,  2002, \mn@doi
  [\PlasmaPhysicsAndControlledFusion] {10.1088/0741-3335/44/7/315}, \href
  {https://ui.adsabs.harvard.edu/abs/2002PPCF...44.1263T} {44, 1263}

\bibitem[\protect\citeauthoryear{{Vaupr{\'e}}, {Hily-Blant}, {Ceccarelli},
  {Dubus}, {Gabici}  \& {Montmerle}}{{Vaupr{\'e}} et~al.}{2014}]{Vaupre2014}
{Vaupr{\'e}} S.,  {Hily-Blant} P.,  {Ceccarelli} C.,  {Dubus} G.,  {Gabici} S.,
    {Montmerle} T.,  2014, \mn@doi [\aap] {10.1051/0004-6361/201424036}, \href
  {http://adsabs.harvard.edu/abs/2014A%26A...568A..50V} {568, A50}

\bibitem[\protect\citeauthoryear{{Voronkov}, {Brooks}, {Sobolev}, {Ellingsen},
  {Ostrovskii}  \& {Caswell}}{{Voronkov} et~al.}{2006}]{Voronkov2006}
{Voronkov} M.~A.,  {Brooks} K.~J.,  {Sobolev} A.~M.,  {Ellingsen} S.~P.,
  {Ostrovskii} A.~B.,   {Caswell} J.~L.,  2006, \mn@doi [\mnras]
  {10.1111/j.1365-2966.2006.11047.x}, \href
  {https://ui.adsabs.harvard.edu/abs/2006MNRAS.373..411V} {373, 411}

\bibitem[\protect\citeauthoryear{{Voronkov}, {Caswell}, {Ellingsen}, {Green}
  \& {Breen}}{{Voronkov} et~al.}{2014}]{Voronkov2014}
{Voronkov} M.~A.,  {Caswell} J.~L.,  {Ellingsen} S.~P.,  {Green} J.~A.,
  {Breen} S.~L.,  2014, \mn@doi [\mnras] {10.1093/mnras/stu116}, \href
  {http://adsabs.harvard.edu/abs/2014MNRAS.439.2584V} {439, 2584}

\bibitem[\protect\citeauthoryear{{Walsh}, {Millar}  \& {Nomura}}{{Walsh}
  et~al.}{2010}]{Walsh2010}
{Walsh} C.,  {Millar} T.~J.,   {Nomura} H.,  2010, \mn@doi [\apj]
  {10.1088/0004-637X/722/2/1607}, \href
  {https://ui.adsabs.harvard.edu/abs/2010ApJ...722.1607W} {722, 1607}

\bibitem[\protect\citeauthoryear{{Wardle}}{{Wardle}}{1999}]{Wardle1999}
{Wardle} M.,  1999, \mn@doi [\apjl] {10.1086/312351}, \href
  {http://adsabs.harvard.edu/abs/1999ApJ...525L.101W} {525, L101}

\bibitem[\protect\citeauthoryear{{Wardle} \& {Yusef-Zadeh}}{{Wardle} \&
  {Yusef-Zadeh}}{2002}]{Wardle2002}
{Wardle} M.,  {Yusef-Zadeh} F.,  2002, \mn@doi [Science]
  {10.1126/science.1068168}, \href
  {http://adsabs.harvard.edu/abs/2002Sci...296.2350W} {296, 2350}

\bibitem[\protect\citeauthoryear{{Yang}, {Xu}, {Choi}, {Ellingsen}, {Sobolev},
  {Chen}, {Li}  \& {Lu}}{{Yang} et~al.}{2020}]{Yang2020}
{Yang} W.,  {Xu} Y.,  {Choi} Y.~K.,  {Ellingsen} S.~P.,  {Sobolev} A.~M.,
  {Chen} X.,  {Li} J.,   {Lu} D.,  2020, \mn@doi [\apjs]
  {10.3847/1538-4365/ab8b5b}, \href
  {https://ui.adsabs.harvard.edu/abs/2020ApJS..248...18Y} {248, 18}

\bibitem[\protect\citeauthoryear{{Yusef-Zadeh}, {Cotton}, {Viti}, {Wardle}  \&
  {Royster}}{{Yusef-Zadeh} et~al.}{2013}]{Yusef-Zadeh2013}
{Yusef-Zadeh} F.,  {Cotton} W.,  {Viti} S.,  {Wardle} M.,   {Royster} M.,
  2013, \mn@doi [\apjl] {10.1088/2041-8205/764/2/L19}, \href
  {http://adsabs.harvard.edu/abs/2013ApJ...764L..19Y} {764, L19}

\bibitem[\protect\citeauthoryear{{Zeng} et~al.,}{{Zeng}
  et~al.}{2020}]{Zeng2020}
{Zeng} S.,  et~al., 2020, \mn@doi [\mnras] {10.1093/mnras/staa2187}, \href
  {https://ui.adsabs.harvard.edu/abs/2020MNRAS.497.4896Z} {497, 4896}

\bibitem[\protect\citeauthoryear{{de Witt}, {Bietenholz}, {Booth}  \&
  {Gaylard}}{{de Witt} et~al.}{2014}]{DeWitt2014}
{de Witt} A.,  {Bietenholz} M.,  {Booth} R.,   {Gaylard} M.,  2014, \mn@doi
  [\mnras] {10.1093/mnras/stt2347}, \href
  {https://ui.adsabs.harvard.edu/abs/2014MNRAS.438.2167D} {438, 2167}

\makeatother
\end{thebibliography}
%\bibliography{interstellar_medium_references,chemistry_references}

\appendix
\section{CH$_3$OH collisional rates}
\label{app_ch3oh_excit}
\begin{figure*}
\includegraphics[width=175mm]{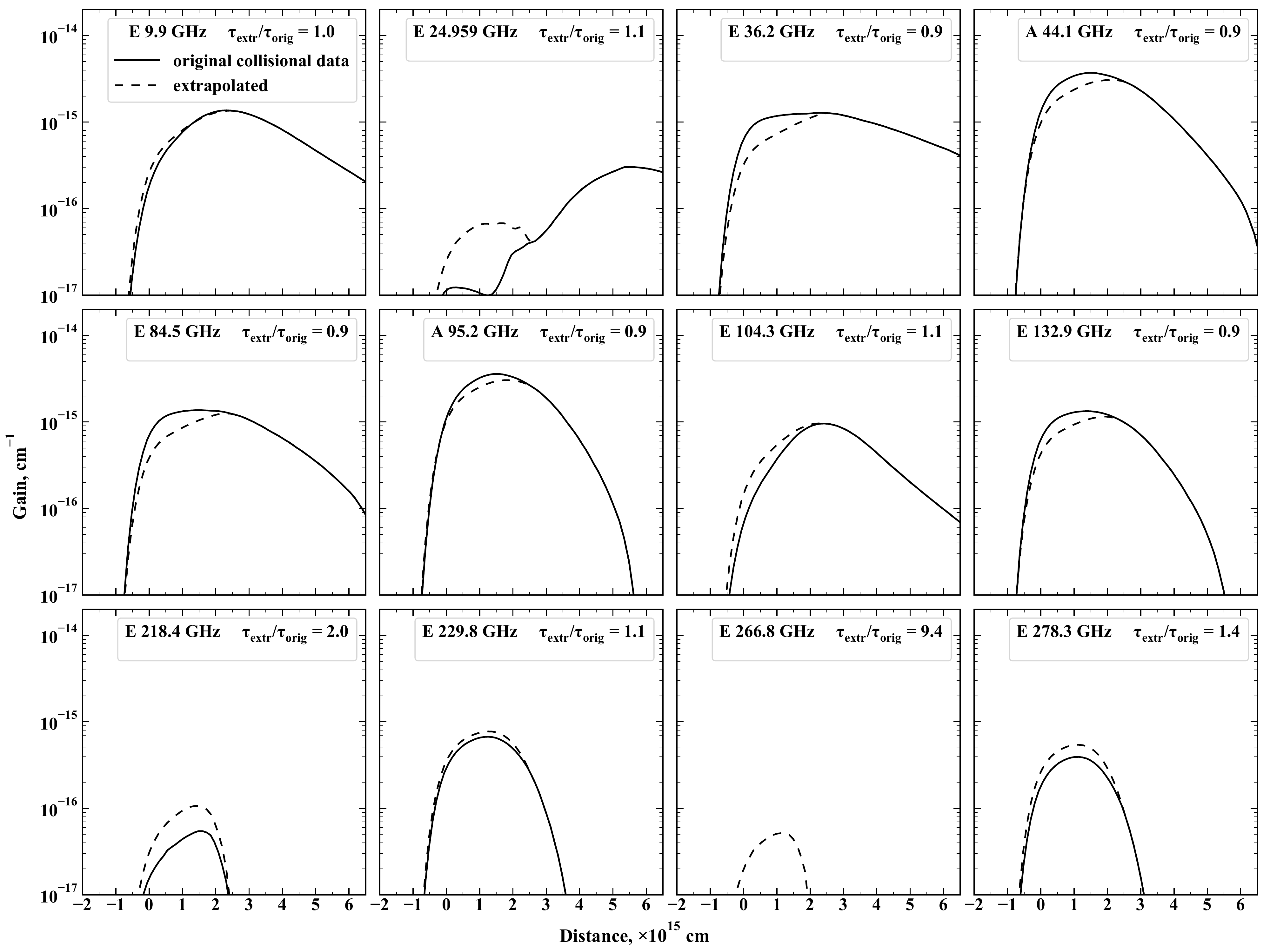}
\caption{Maser gain in methanol transitions as a function of the distance in the shock wave. The solid line corresponds to the calculations without extrapolation of collisional rate coefficients. The dashed line corresponds to the calculations in which the extrapolation of rate coefficients is employed at high gas temperatures. The ratio of optical depths at the line centre calculated using original and extrapolated collisional rate coefficients is indicated on each plot. A high aspect ratio is adopted in the calculations of optical depths, $1/\mu = 10$. The parameters of the shock model are $\zeta_\mathrm{H_2} = 3 \times 10^{-17}$~s$^{-1}$, $n_\text{H,tot} = 2 \times 10^5$~cm$^{-3}$, and $u_\text{s} = 17.5$~km~s$^{-1}$.}
\label{fig_app1}
\end{figure*}

Methanol is ejected to the gas via the sputtering of grain mantles at the shock front. The contribution of the hot gas near the shock peak to the total methanol line emission may be substantial. In order to estimate the effect of high temperature collisional data, the calculations have been done with the collisional rate coefficients that are assumed to increase as $T^{1/2}$ at high gas temperatures. This assumption implies that the rate coefficients depend on temperature through the collision velocity, while the de-excitation cross sections are assumed constant \citep{Schoier2005}. We compare two simulations with and without the extrapolation of collisional data in Fig.~\ref{fig_app1}. The difference in optical depths at the line centre calculated using original and extrapolated collisional rate coefficients is small for strong masers. However, this difference may be substantial for weak lines, in particular for E type methanol transitions of the $J_2 \to (J-1)_1$ series at 218.4 and 266.8~GHz, see Fig.~\ref{fig_app1}. 

The energy levels belonging to the ground and two lowest torsionally excited states are taken into account in our calculations. For the model presented in Fig.~\ref{fig_app1}, most of the methanol molecules populate the ground state. For example, the maximal fractions of methanol molecules of each type in the first and second torsionally excited states are equal to about $10^{-6}$ and $10^{-7}$, respectively. The high rates of spontaneous radiative de-excitation of methanol ensure that the population densities of high-lying energy levels remain small in the absence of an external radiation field \citep{Flower2010b}.

\label{lastpage}
\end{document}